# Twist-configured moiré–moiré reconstruction governs diverse commensurate double-moiré phases in twisted bilayer graphene on h-BN

*Yuta Seo[1*], Naoto Nakatsuji[2*], Jimpei Kawase[1], Naoto Hishida[1], Kenji Watanabe[3], Takashi Taniguchi[4], Takuto Kawakami[5], Mikito Koshino[6] and Tomoki Machida[1]*

*These authors contributed equally to this work.

[1] Institute of Industrial Science, The University of Tokyo, 4-6-1 Komaba, Meguro, Tokyo 153-8505, Japan

[2] Department of Physics and Astronomy, Stony Brook University, Stony Brook, New York 11794, USA

[3] Research Center for Electronic and Optical Materials, National Institute for Materials Science, 1-1 Namiki, Tsukuba 305-0044, Japan

[4] Research Center for Materials Nanoarchitectonics, National Institute for Materials Science, 1-1 Namiki, Tsukuba 305-0044, Japan

[5] Center for Integrated Science and Humanities, Fukushima Medical University, Fukushima 960-1295, Japan

[6] Institute for Solid State Physics, The University of Tokyo, Kashiwa 277-8581, Japan

Mail: yu-seo@iis.u-tokyo.ac.jp, naoto.nakatsuji@stonybrook.edu, koshino@issp.u-tokyo.ac.jp, tmachida@iis.u-tokyo.ac.jp

**Abstract**

The coexistence of multiple moiré lattices in van der Waals heterostructures raises a fundamental question: how do distinct moiré patterns interact and reconstruct? Here, we investigate twisted bilayer graphene (tBG) on hexagonal boron nitride (h-BN), where tBG and graphene/h-BN moiré structures coexist, using conductive atomic force microscopy combined with continuum-model simulations. We show that reconstruction between these moiré lattices—moiré–moiré reconstruction—manifests across multiple length scales, giving rise to diverse commensurate double-moiré phases. Locally, the stacking registry between the two moiré lattices is uniquely selected by the global twist configuration (helical or alternate), mediated by rotational relaxation of the shared graphene layer. This registry, together with twist angle and strain, governs commensurate domains from $C_{3z}$-symmetric to strained symmetry-modified structures. These results establish moiré–moiré reconstruction as a general framework for engineering structural and electronic order—including theoretically predicted topological flat bands below the magic angle—in multilayer moiré materials.

## Main

## Introduction

Moiré superlattices, long-wavelength interference patterns generated by stacking atomically thin crystals with a small lattice mismatch or relative twist, have emerged as a powerful platform for engineering electronic structures in two-dimensional materials[1–16]. In canonical single-interface moiré systems such as twisted bilayer graphene (tBG)[17–19] and graphene on hexagonal boron nitride (Gr/h-BN)[20,21], atomic-scale structural relaxation reorganizes the moiré pattern into commensurate domains and domain-wall networks, profoundly modifying the electronic structure and giving rise to a variety of emergent phenomena[17,19,20]. Structural reconstruction is therefore recognized as a central ingredient of moiré physics, linking atomic-scale energetics to mesoscale order and quantum electronic properties.

As moiré research advances toward increasingly complex multilayer architectures, attention has shifted to multi-moiré systems in which multiple moiré patterns from different interfaces coexist within the same heterostructure. Unlike single-interface systems, distinct moiré superlattices in a multilayer structure are coupled through shared atomic layers, such that lattice relaxation associated with one interface can directly influence that of another. This coupling introduces a new class of collective structural phenomena beyond conventional single-interface moiré physics and raises a fundamental question: how do different moiré superlattices interact through lattice relaxation, and what principles govern the resulting real-space organization?

Initial progress has been made in several classes of multi-moiré systems. In homo-type structures, such as twisted trilayer graphene, the coexistence of multiple moiré interfaces leads to complex reconstructions[22–25]. Hetero-type systems provide an even richer setting by combining moiré superlattices of distinct microscopic origin. In particular, tBG aligned with h-BN hosts two qualitatively different moiré patterns simultaneously: a triangular moiré lattice associated with the graphene–graphene interface and a hexagonal moiré lattice generated by the graphene–h-BN interface. Recent experiments have revealed moiré self-alignment and local commensurate structures[26–28] in limited regions of parameter space, suggesting that the two moiré patterns may become locally synchronized through structural relaxation. However, it remains unclear whether such commensuration is a special feature of particular twist-angle combinations or a general consequence of moiré–moiré interaction. More broadly, the roles of global twist configuration (helical or alternate twisting), twist angle, and strain in determining local stacking registry and domain formation remain largely unexplored.

Here we show that moiré–moiré reconstruction—the collective lattice relaxation between coexisting moiré superlattices—acts as a general organizing principle in tBG/h-BN hetero-trilayers, shaping structural and electronic order across multiple length scales. Combining conductive atomic force microscopy (C-AFM) with continuum-model simulations, we find that the global twist configuration (helical or alternate) deterministically selects the local spatial registry between the two moiré lattices through a local rotation matching mechanism in the shared graphene layer. In helical configuration, vertices of the triangular tBG moiré lattice align with the centers of hexagonal Gr/h-BN moiré domains, whereas in alternate configuration they align with the domain vertices of the Gr/h-BN moiré. This registry, together with the twist angle and strain, governs the formation of a rich family of mesoscale commensurate double-moiré phases—from $C_{3z}$-symmetric to strain-modified symmetry-broken structures. Remarkably, strain does not destroy commensuration; instead, it gives rise to symmetry-broken yet registry-preserved commensurate domains, demonstrating that the local rotation matching mechanism operates as a fundamental organizing principle regardless of strain conditions. At larger length scales, moiré–moiré reconstruction promotes well-ordered sub-micrometer domains with boundaries that exhibit collective sliding dynamics. Theoretically, the twist-configuration-dependent registries stabilize topological flat bands with distinct Chern numbers even below the magic angle. These results establish moiré–moiré reconstruction based on the local rotation matching mechanism as a predictive and general framework for engineering structural and electronic order in multilayer moiré systems—opening a design space inaccessible to single-moiré physics.

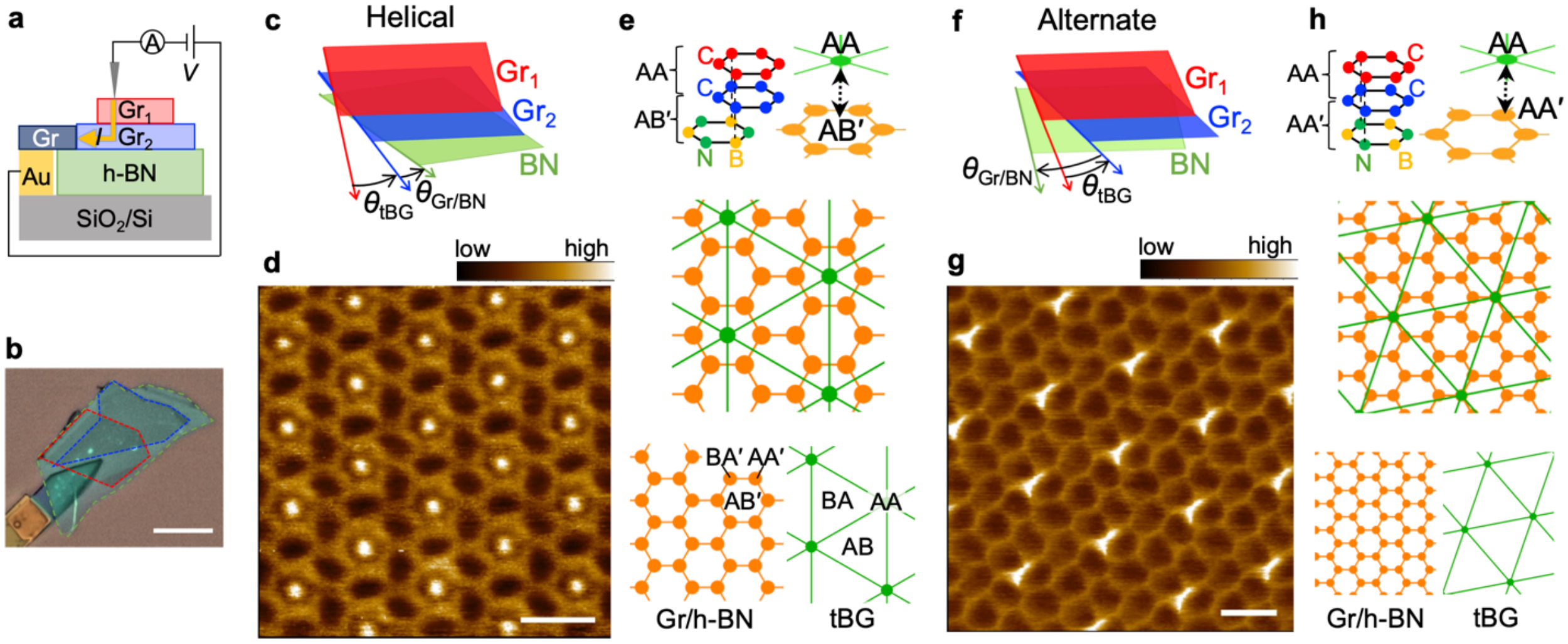

**Figure 1. Twist-configuration-dependent moiré–moiré stacking registry in tBG/h-BN hetero-trilayers. a,** Schematic of the conductive atomic force microscopy (C-AFM) measurement on the tBG/h-BN hetero-trilayer. **b,** Optical image of a helical-twisting device. Scale bar, 10 μm. **c,** Schematic of the helical-twisting geometry. **d,** C-AFM image of the commensurate helical double-moiré lattice [ $(\theta_{\mathrm{tBG}}, \theta_{\mathrm{Gr/BN}}) = (0.60°, 0.62°)$; $L_{\mathrm{tBG}}: L_{\mathrm{Gr/BN}} = 2:1$ ]. **e,** Schematic of the reconstructed helical double-moiré structure. Top: atomic-scale illustration of the observed local stacking registry, showing tBG AA stacking—where carbon atoms of both graphene layers are vertically aligned—positioned on the AB′ stacking of the Gr/h-BN moiré—where graphene carbon atoms sit above boron atoms. Middle: schematic of commensurate double-moiré structure corresponding to Fig. 1d. Bottom: decomposition into isolated tBG and Gr/h-BN moiré structures. In tBG, AB and BA denote Bernal stacking, with AA stacking at domain vertices. In Gr/h-BN, AA′, AB′, and BA′ denote high-symmetry stacking configurations, where graphene carbon atoms sit above both boron and nitrogen atoms (AA′), above boron atoms (AB′), or above nitrogen atoms (BA′). **f,** Schematic of the alternate-twisting geometry. **g,** C-AFM image of the commensurate alternate double-moiré lattice [ $(\theta_{\mathrm{tBG}}, \theta_{\mathrm{Gr/BN}}) = (0.39°, -0.97°)$; $L_{\mathrm{tBG}}: L_{\mathrm{Gr/BN}} = \sqrt{7}:1$ ]. **h,** Schematic of the reconstructed alternate double-moiré structure. Top: atomic-scale illustration of the observed local stacking registry, showing tBG AA stacking positioned on AA′ stacking of the Gr/h-BN moiré. Middle: schematic of commensurate double-moiré structure corresponding to Fig. 1g. Bottom: decomposition into isolated tBG and Gr/h-BN moiré structures (stacking definitions as in panel e). Scale bars of Fig. 1d,g, 20 nm.

## Results and Discussion

Figure 1a illustrates the C-AFM setup used to image the moiré structures of tBG/h-BN samples (Fig. 1b). In contact mode, the tip scans the sample surface under an applied bias, generating a current that reflects stacking-dependent local electronic properties, enabling visualization of moiré superlattices[29–37]. We fabricated tBG/h-BN stacks in helical (successive twists in the same direction; Fig. 1c) and alternate (successive twists in alternating directions; Fig. 1f) configurations via a dry-transfer method (see Methods).

Fig. 1d,g present representative C-AFM current images for the helical and alternate configurations, respectively. In both images, two distinct features are simultaneously resolved: bright dots forming a triangular periodic pattern and a periodically distorted hexagonal network. In the helical configuration (Fig. 1d), the bright dots consistently register within the hexagonal cells, whereas in the alternate configuration (Fig. 1g), they align with the vertices of the hexagonal network. As established in prior work, isolated tBG reconstructs into triangular AB and BA domains with AA stacking at their vertices[17–19], while Gr/h-BN forms a hexagonal AB′ domain network with AA′ and BA′ at their vertices[20,21] (see Fig. 1e,h and captions for stacking notation). The observed bright dots and hexagonal network are thus identified as tBG AA domains and the Gr/h-BN moiré, respectively. The observed moiré pattern (Fig. 1d,g) therefore reveals that the spatial registry between the tBG and Gr/h-BN moirés is uniquely determined by the twist configuration. In the helical configuration, the vertices of the triangular tBG moiré lattice (AA) align with the centers of the Gr/h-BN moiré cells (AB′) (Fig. 1e), whereas in the alternate configuration they align with the vertices of the Gr/h-BN moiré (AA′/BA′) (Fig. 1h). Furthermore, in both configurations the hexagonal Gr/h-BN pattern exhibits distortions synchronized with the tBG moiré, indicating strong mutual coupling between the two moiré lattices.

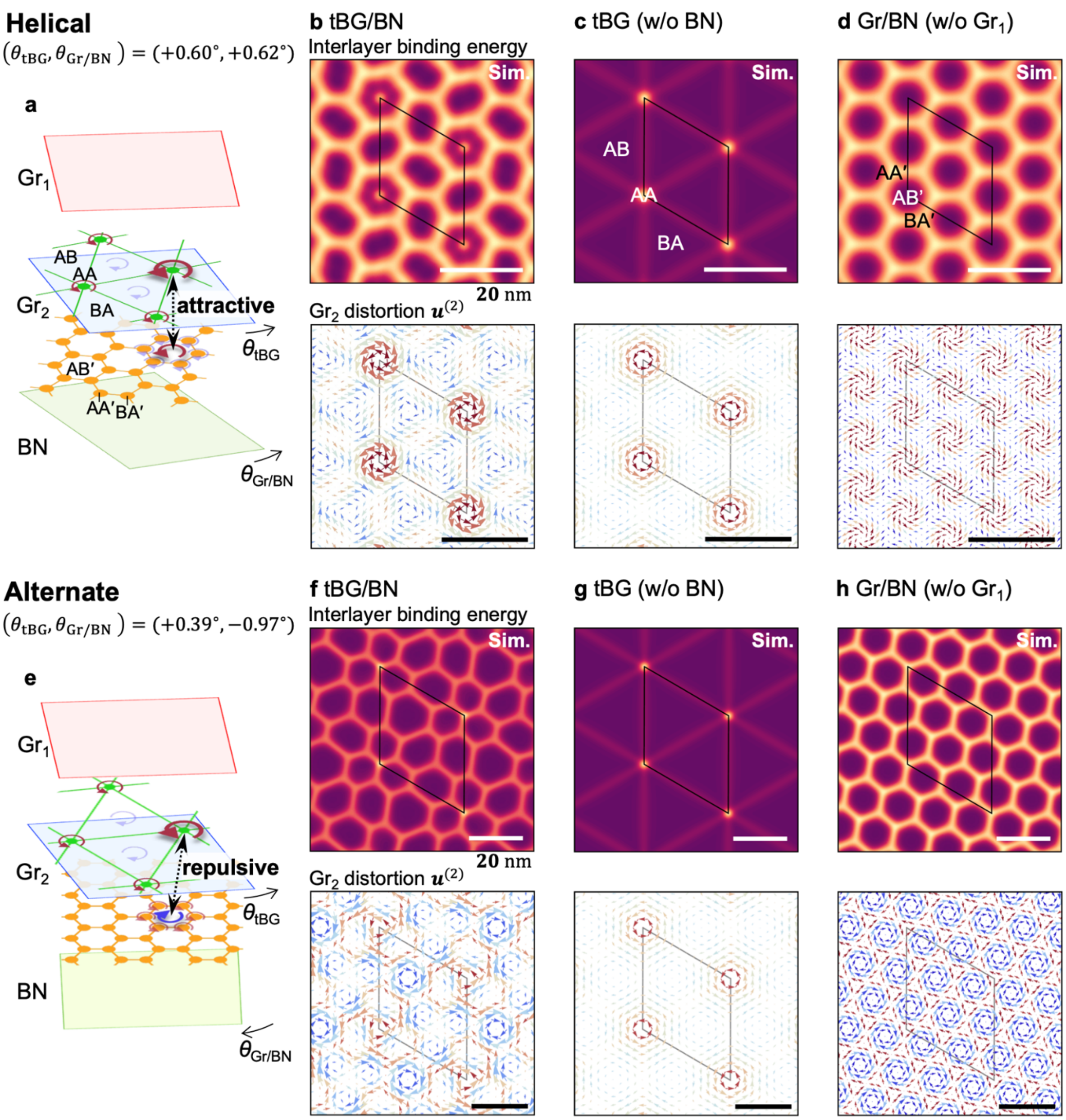

**Figure 2. Lattice-relaxation-driven local rotation matching mechanism governing moiré–moiré stacking registry. a,** Schematic illustrating the local rotation matching in the helical-twisting case. Red, blue, and green sheets represent $Gr_1$, $Gr_2$ and h-BN, respectively. Red and blue arrows indicate the local rotational response of the shared graphene ($Gr_2$) in tBG and Gr/h-BN moirés. **b,** Real-space binding-energy landscape (top), and displacement field of $Gr_2$ (bottom) for the relaxed tBG/h-BN with ($\theta_{tBG}$, $\theta_{Gr/BN}$) = (0.60°, 0.62°). In the bottom figure, arrow colors represent the rotational components at each position - blue and red denote clockwise and counterclockwise rotation, respectively. Arrow size indicates the magnitude of the displacement vectors. **c-d,** Corresponding results for isolated (c) tBG and (d) Gr/h-BN with twist angles matching those in **b**. **e-h,** Similar figures for the alternate twist tBG/h-BN with ($\theta_{tBG}$, $\theta_{Gr/BN}$) = (0.39°, -0.97°). Scale bars, 20 nm. These results establish local rotation matching as the deterministic mechanism governing moiré–moiré registry.

Here, using continuum-model simulations, we demonstrate that the twist-configuration-dependent stacking registry is governed by the coupling between the tBG and Gr/h-BN moiré lattices mediated by the rotational relaxation of the shared graphene layer $Gr_2$: the two moiré lattices are arranged such that the directions of the local rotations induced in $Gr_2$ by each moiré for their individual relaxation match each other (Fig. 2a,e).

The top panel of Fig. 2b shows the calculated local interlayer binding-energy landscape for the helical-twisting case, which reveals the local stacking structures responsible for the C-AFM current contrast. Remarkably, this theoretical result is in excellent agreement with the experimental C-AFM patterns, reproducing not only the characteristic stacking registry but also the detailed distortion patterns of the Gr/h-BN hexagonal network synchronized with the tBG moiré.

To identify the microscopic origin, we analyze the displacement vector of the shared graphene layer, $\mathbf{u}^{(2)}$, defined as the in-plane deviation of atomic positions from the rigid lattice (see SI Section II). The bottom panel of Fig. 2b shows the displacement field $\mathbf{u}^{(2)}(\boldsymbol{r})$ for the helical configuration. The arrows represent the local displacement vectors, while the color indicates the rotational component $\Omega(\boldsymbol{r}) = (\partial_x u_y - \partial_y u_x)/2$, with blue and red corresponding to clockwise and counterclockwise local rotations, respectively. In single-moiré systems, lattice relaxation locally reduces the twist angle in energetically favorable stacking regions to expand them, while increasing the twist angle near unfavorable regions to suppress them[17,19,38]. In the tBG/h-BN double moiré lattice, the rotational components in the shared graphene layer are expected to reflect lattice relaxation at both tBG and Gr/h-BN interfaces; we therefore decompose the pattern into its tBG and Gr/h-BN contributions (Fig. 2c,d bottom panels).

When considering only the tBG moiré component of the helical configuration (Fig. 2a), lattice relaxation drives a counterclockwise rotation of $Gr_2$ near the AA sites (Fig. 2c). In the Gr/h-BN moiré component, $Gr_2$ rotates counterclockwise near AB′ sites but clockwise near AA′/BA′ sites (Fig. 2d). When these lattices coexist, an energetically stable registry is achieved only where these rotational tendencies are mutually reinforcing. As shown in Fig. 2c,d, the counterclockwise rotations preferred by both tBG AA and Gr/h-BN AB′ sites drive $Gr_2$ in the same direction, rendering their vertical alignment—the AA on AB′ registry—energetically favorable in the helical configuration.

In the alternate-twisting case (Fig. 2e–h), the reversed twist direction of the h-BN layer relative to the helical case inverts the rotational field of the Gr/h-BN moiré (Fig. 2h). Consequently, $Gr_2$ now prefers counterclockwise rotation near the AA′/BA′ regions instead. To maintain cooperative relaxation, the tBG AA domains adaptively align with these Gr/h-BN AA′/BA′ regions (Fig. 2g,h).

The local rotational displacement of $Gr_2$ that underlies the rotation matching mechanism simultaneously modulates the local twist angle at the Gr/h-BN interface. In the helical configuration, this produces periodic distortions of the Gr/h-BN hexagonal network synchronized with the tBG moiré (Fig. 2b). In the alternate configuration, near tBG AA sites, the local rotation of $Gr_2$ increases the twist angle at the Gr/h-BN interface and shortens its local moiré period, whereas near AB/BA sites it decreases the twist angle and enlarges the period (Fig. 2f), producing a spatially modulated Gr/h-BN periodicity that directly tracks the tBG superlattice. These results establish local rotation matching as the deterministic mechanism governing moiré–moiré registry.

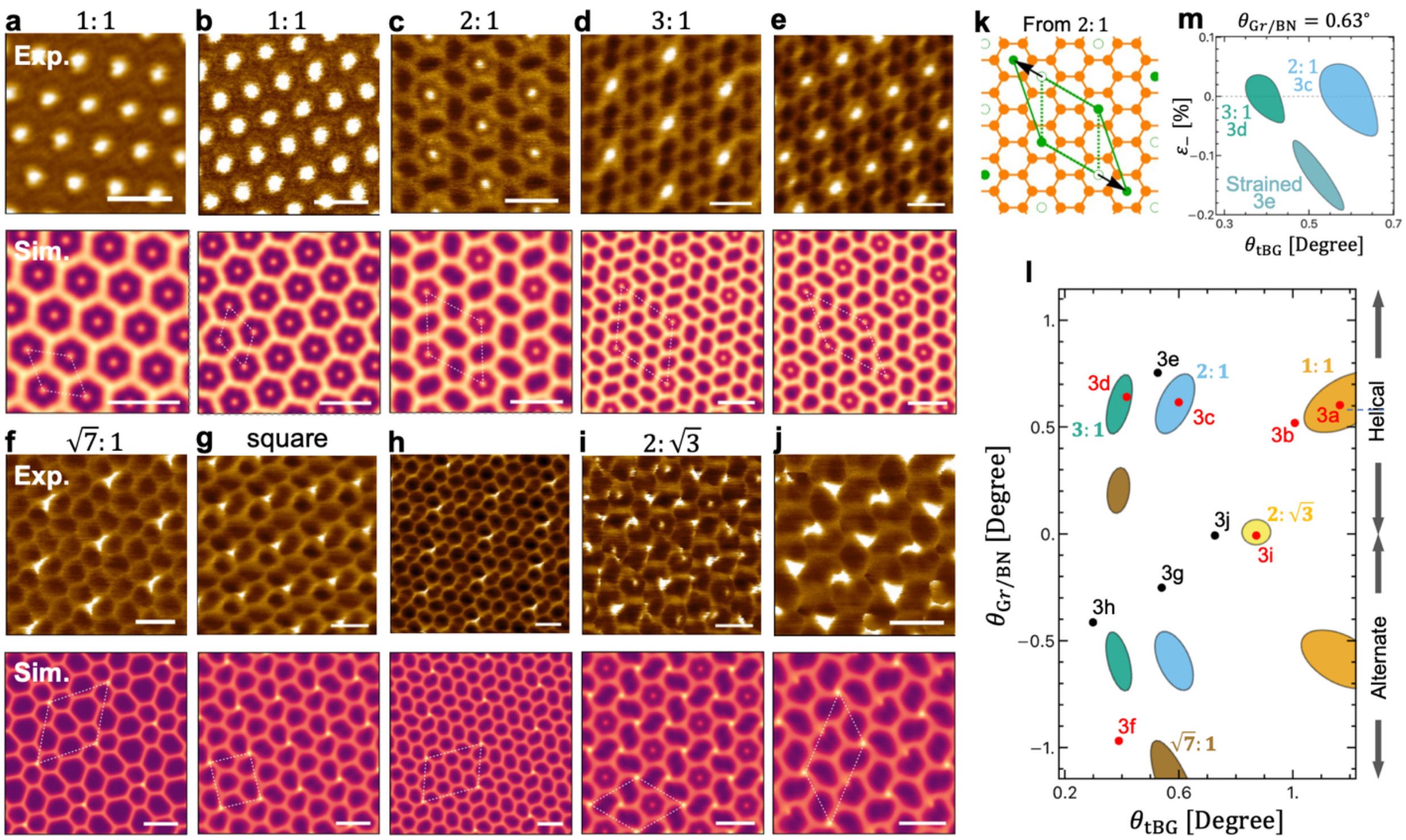

**Figure 3. Structural diversity of the helical and alternate double moiré lattices. a–e**, Experimental C-AFM images (top) and simulated structures (bottom) for helical-twisting samples. **a-d,** Helical $C_{3z}$-symmetric commensurate structures with $L_{tBG}$: $L_{Gr/BN}$ =1:1 (a, b), 2:1 (c), and 3:1 (d), respectively. **e,** Strained ($C_{3z}$ symmetry breaking) commensurate structure, still forming large scale domain (see also Fig. 4a). **f–j**, Corresponding panels for alternate-twisting samples. **f,** $C_{3z}$-symmetric commensurate structure with $L_{tBG}$: $L_{Gr/BN}$ =$\sqrt{7}$: 1. **g,** Strained commensurate structure with square-like lattice, stabilized over a large area (see also Fig. 4b). **h,** Strained commensurate configuration with a distinct pattern. **i,** $C_{3z}$-symmetric commensurate structure with $L_{tBG}$: $L_{Gr/BN}$ =2: $\sqrt{3}$, with $\theta_{Gr/BN}$~0°. **j,** Strained commensurate configuration having $\theta_{Gr/BN}$~0°. Scale bars, 20 nm. **k,** Schematic illustrations of the configuration transition from the $C_{3z}$-symmetric states with $L_{tBG}$: $L_{Gr/BN}$ = 2:1. **l,** Commensurate-domain phase diagram as a function of $\theta_{tBG}$ and $\theta_{Gr/BN}$. The color areas represent the twist angle regions capable of forming commensurate domains [Orange-1:1, sky blue-2:1, bluish green-3:1, brown-$\sqrt{7}$:1, yellow-2: $\sqrt{3}$]. The markers indicate the twist conditions of the panels (a)-(j), where the red and black dots represent the nearly $C_{3z}$ symmetric and strained cases, respectively. **m,** Commensurate configuration phase diagram as a function of strain $\varepsilon_-$ and $\theta_{tBG}$.

Moiré–moiré reconstruction gives rise to a rich set of commensurate double-moiré structures across various twist angles and strain conditions. Fig. 3a–j presents experimental C-AFM images (top panels) and corresponding simulated structures (bottom panels) from a systematic survey, covering both $C_{3z}$-symmetric and strained symmetry-modified commensurate structures in helical (Fig. 3a-e) and alternate (Fig. 3f-j) configurations. In all cases, the local stacking registry established in Figs. 1 and 2 is preserved. Moreover, the local stacking registry repeats periodically across extended spatial regions, which we refer to as commensurate domains.

In the helical configuration, $C_{3z}$-symmetric commensurate domains with quantized integer period ratios $L_{\mathrm{tBG}} / L_{\mathrm{Gr/BN}} = 1$, 2, and 3 (Fig. 3a–d) are observed, consistent with prior STM study[28]. In the alternate configuration, a nearly $C_{3z}$-symmetric commensurate domain ($L_{\mathrm{tBG}} / L_{\mathrm{Gr/BN}} = \sqrt{7}$) is observed (Fig. 3f). In these commensurate domains, the sites with the twist-configured stacking registry form a $C_{3z}$-symmetric pattern, reflecting the three-fold rotational symmetry of the constituent moiré lattices.

In addition to these $C_{3z}$-symmetric phases, commensurate domains with broken $C_{3z}$ symmetry are also observed in experiment (Fig. 3e,g,h top panels). In these structures, the tBG moiré lattice is significantly distorted, breaking the $C_{3z}$ symmetry of the commensurate domain, yet the twist-configured stacking registry is preserved. These structures are quantitatively reproduced by theoretical simulations (Fig. 3e, g, h bottom panels) incorporating hetero-strain at the tBG interface and homo-strain at the graphene/h-BN interface (see SI Section IV for details). In general, the tBG and Gr/h-BN moiré lattices differ in both their periods and orientations, making them incommensurate and giving rise to a moiré-of-moiré periodicity that would modulate the structure across the sample; sustaining a uniform commensurate structure over large areas would therefore be extremely unlikely (see SI Section III for details). Yet such commensurate domains are observed over large areas, demonstrating that moiré–moiré reconstruction stabilizes commensurate domains regardless of whether strain is present. The quantitative agreement between experiment and simulation further validates the local rotation matching mechanism as a predictive framework for commensurate domain formation.

Building on this framework, we construct a commensurate-domain phase diagram as a function of $\theta_{tBG}$ and $\theta_{Gr/BN}$ (Fig. 3l). The colored regions in Fig. 3l indicate twist-angle combinations ($\theta_{tBG}$, $\theta_{Gr/BN}$) under which commensurate domains are predicted to form through moiré–moiré reconstruction. Without reconstruction, exact commensurability between the two moiré lattices would require simultaneous matching of both their periods and orientations, achievable only at discrete combinations of ($\theta_{tBG}$, $\theta_{Gr/BN}$)[39–42]. Moiré–moiré reconstruction extends this to a finite window of twist angles within which the two moiré lattices are sufficiently close to exact commensurability that a small lattice distortion can lock them into a commensurate domain. Fig. 3l shows the theoretically predicted window within which the residual mismatch from exact commensurability, $|\Delta \boldsymbol{L}|$, is small enough to satisfy $|\Delta \boldsymbol{L}|/\max(|\boldsymbol{L}_{\mathrm{tBG}}|, |\boldsymbol{L}_{\mathrm{Gr/BN}}|) < 0.1$, allowing reconstruction to stabilize a commensurate domain[23] (see SI Section II for the precise definition of $\Delta \boldsymbol{L}$). Within this regime, moiré–moiré reconstruction reduces the total energy by selecting and expanding energetically favorable stacking registries, thereby stabilizing commensurate domains. The observed $C_{3z}$-symmetric structures (Fig. 3a–d, f) are found within the predicted regions, validating the phase diagram. The $\sqrt{7}$:1 structure in Fig. 3f lies slightly outside the predicted commensurate window, likely due to strain effects.

The strain-dependent phase diagram (Fig. 3m) is constructed by applying hetero-strain to the tBG interface and homo-strain to the graphene/h-BN interface, and identifying the range of strain values at given twist angles for which commensurate domains are predicted to form, using the same near-commensurability condition as employed for Fig. 3l (see SI Section IV for details). This analysis reveals that strained commensurate domains are indeed stabilized through the moiré–moiré reconstruction within finite strain windows, with the stacking registry preserved despite the breaking of $C_{3z}$ symmetry. The strained structures (Fig. 3e, g, h) can be categorized as symmetry-modified counterparts of neighboring $C_{3z}$-symmetric phases. For example, the strained structure in Fig. 3e is found near the 2:1 and 3:1 commensurate regions in the phase diagram (Fig. 3l,m), suggesting that these two phases are separated by only a small strain. This is consistent with the experimental result in Fig. 4a, where the strained (Fig. 3e) and 2:1 commensurate (Fig. 3c) domains coexist in close proximity. Similarly, the strained configurations shown in Fig. 3g and 3h can be associated with the neighboring $C_{3z}$-symmetric phases (Fig. S4). While the strained phases are stabilized when strain introduced during fabrication shifts the system toward a strained commensurate window, the $C_{3z}$-symmetric phases are expected to represent the lowest-energy commensurate configurations at a given twist angle. Because the Gr/h-BN moiré template confines tBG AA domains to discrete sites, the system undergoes discontinuous structural transitions between commensurate phases under strain (Fig. 3k, S4)—in contrast to isolated tBG, where moiré patterns evolve continuously.

Having established the roles of twist angle and strain, we now consider the limits of commensurate domain formation. Near $\theta_{\mathrm{Gr/BN}} \approx 0°$, lattice relaxation at the Gr/h-BN interface is dominated by local expansion and contraction rather than rotation, suppressing the local rotation matching mechanism and hindering large-area commensurate domain formation (Fig. 3i top panel; see SI Section V). When commensurate conditions are not met, the local stacking registry is preserved but long-range periodic order is absent (Fig. S5). Taken together, the global twist configuration, twist angle, and strain govern commensurate domain formation in tBG/h-BN.

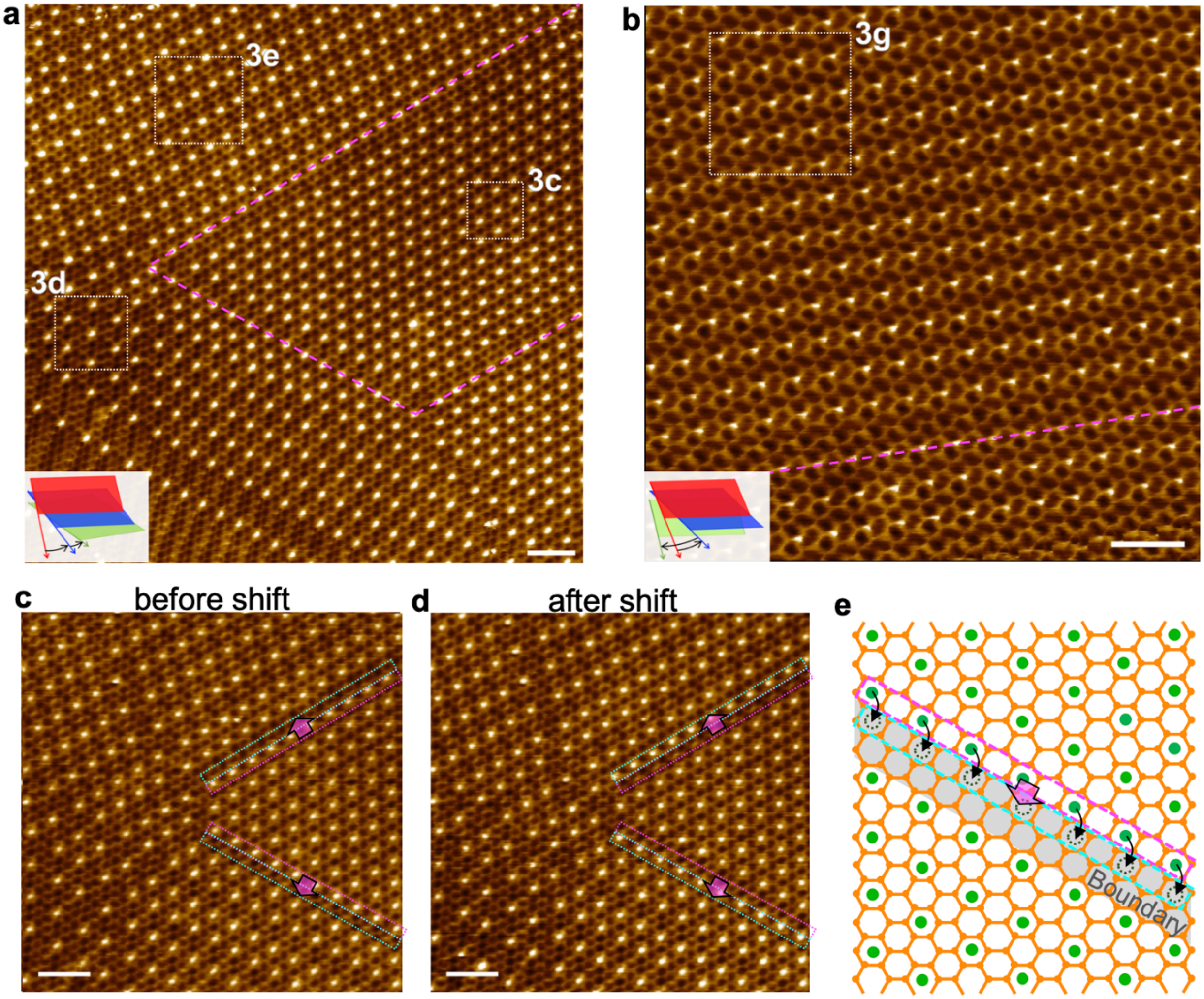

**Figure 4. Mesoscopic order and collective boundary dynamics in helical and alternate tBG/h-BN double-moiré lattices. a, b,** Large-area C-AFM current maps of helical (a) and alternate (b) configurations, showing well-defined sub-micrometer-scale domains separated by long, nearly straight boundaries intersecting at multiples of 60°. Labels indicate regions exhibiting the commensurate structures classified in the corresponding panels of Fig. 3. **c, d,** Sequential C-AFM images of a representative domain-boundary region in (a), acquired before (c) and after (d) a shift, with a 90 nN scan in between. Arrows indicate the coordinated displacement of a chain of tBG AA domains along the domain boundary. **e,** Schematic illustration of the collective sliding of the tBG AA domains along a domain boundary, corresponding to the behavior shown in (c) and (d). Scale bars, 50 nm.

Large-area C-AFM images reveal well-ordered mesoscopic domain structures separated by long, nearly straight boundaries, highlighted by dashed lines in Fig. 4a,b. Within each domain, the tBG moiré forms a highly ordered lattice, whereas the periodic order of the double-moiré structure is disturbed across the domain boundaries (see SI Section VI for a detailed discussion of domain boundary morphology). This large-scale ordering may be attributed to the Gr/h-BN moiré, whose moiré period is relatively insensitive to the twist angle and thus remains spatially uniform across the sample, acting as a periodic template that organizes the tBG moiré into well-ordered domains.

In addition to this mesoscopic structural organization, the domain boundaries exhibit a distinct dynamic signature. During C-AFM scans, chains of tBG AA domains located along selected boundaries undergo collective sliding parallel to the boundary direction (Fig. 4c–e). Rather than moving independently, these AA domains shift coherently as a quasi-one-dimensional chain without altering the surrounding domain structure.

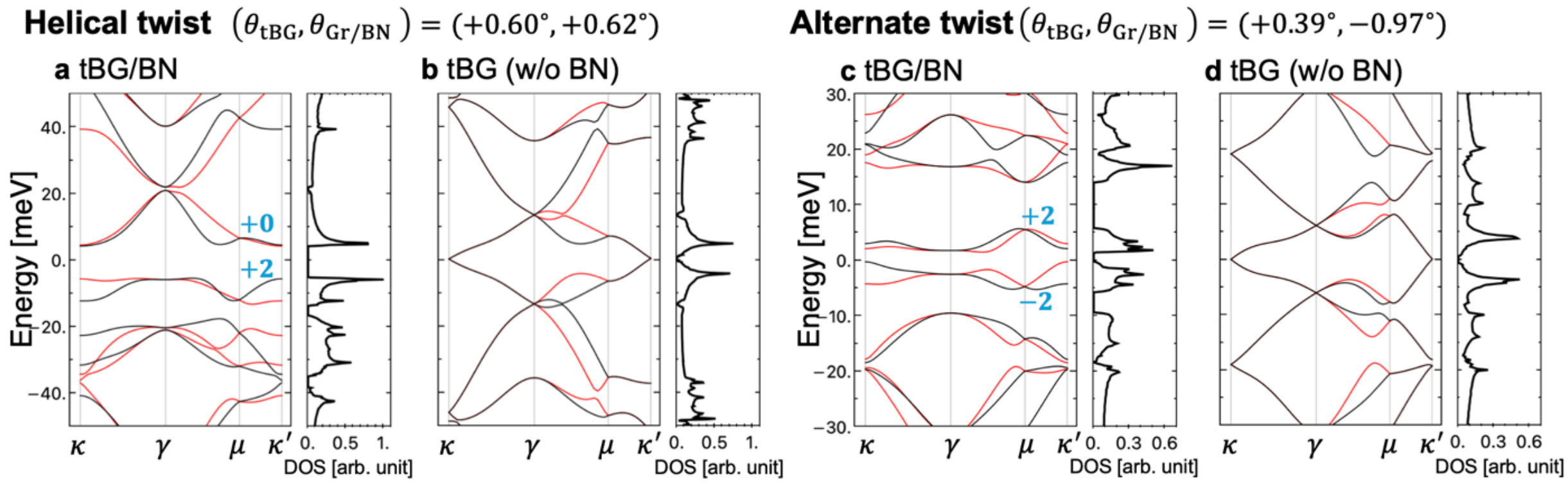

**Figure 5. Electronic consequences of moiré–moiré stacking. a,** Electronic band structure and density of states for the relaxed helical configuration at ($\theta_{tBG}$, $\theta_{Gr/BN}$) = (0.60°, 0.62°), corresponding to the 2:1 commensurate structure in Fig. 3c. **b,** Band structure of the corresponding relaxed tBG without the Gr/h-BN moiré, for comparison. **c,** Band structure and DOS for the relaxed alternate configuration at ($\theta_{tBG}$, $\theta_{Gr/BN}$) = (0.39°, -0.97°), corresponding to the $\sqrt{7}$: 1 commensurate structure in Fig. 3f. **d,** corresponding relaxed-tBG calculation. Black and red curves indicate K and K′ valleys, respectively; blue labels show Chern numbers of the K-valley bands. These calculations reveal that moiré–moiré stacking modifies the tBG minibands, opening a gap at charge neutrality and enabling flat-band formation well below the magic angle.

Finally, we examine the electronic consequences of the structural registries. To account for the different spatial alignment between the tBG and Gr/h-BN moirés in the helical and alternate configurations, we compute the band structure using a model that treats the Gr/h-BN moiré as a spatially varying potential. Figure 5a–d presents the calculated band structures and density of states (DOS) for both helical- and alternate-twisting configurations, together with relaxed tBG at the same twist angles without the Gr/h-BN moiré for comparison. Although the low-energy minibands originate primarily from the tBG component, the Gr/h-BN moiré potential introduces an essential perturbation that opens gaps at both the κ/κ′ points and the γ point, leading to isolated topological flat bands at twist angles well below the magic angle—a regime in which tBG alone does not exhibit such band isolation[17]. In the alternate configuration (Fig. 5c,d), topological flat bands with higher valley Chern numbers ($|C| = 2$) are obtained. Similar topological flat bands also appear in the helical configuration (Fig. 5a,b), but the detailed miniband dispersions and valley Chern allocations differ. This twist-configuration-dependent electronic band structure is explicitly demonstrated by the comparison of helical and alternate 2:1 configurations at the same tBG twist angle in Fig. S7 (SI Section VII). The global twist configuration therefore can serve as a parameter for controlling miniband topology in double-moiré systems.

**Conclusion**

We have shown that moiré–moiré reconstruction governs structural order across multiple length scales, giving rise to diverse commensurate double-moiré structures. At the local scale, the global twist configuration (helical or alternate) uniquely selects the moiré–moiré stacking registry through the local rotation matching mechanism: the two moiré lattices adopt the configuration in which the directions of the local rotations induced in the shared graphene layer by each interface match each other. Beyond this local registry, moiré–moiré reconstruction drives the formation of extended commensurate domains whose structure is governed by three parameters—the global twist configuration, the twist angle, and strain, giving rise to the $C_{3z}$-symmetric phases at quantized period ratios and symmetry-modified strained phases within adjacent strain windows with the moiré–moiré stacking registry preserved. At the mesoscopic scale, the Gr/h-BN template promotes the formation of well-ordered sub-micrometer commensurate domains whose boundaries exhibit collective sliding dynamics.

Beyond structural organization, the twist-configured moiré–moiré registry has direct electronic consequences. The Gr/h-BN moiré potential introduces essential perturbations that open gaps and stabilize topological flat bands with twist-configuration-dependent Chern numbers, establishing the global twist configuration as a crucial parameter for engineering miniband topology in double-moiré systems. These predictions are experimentally accessible via scanning tunneling spectroscopy, which would reveal the predicted gap openings and flat bands, and Landau level spectroscopy, which would probe the twist-configured valley Chern numbers.

Since rotational relaxation is a generic feature of van der Waals moiré systems, the local rotation matching mechanism identified here is expected to apply broadly across diverse material families and multilayer moiré materials with increasing numbers of moiré interfaces. As van der Waals heterostructures of increasing complexity become experimentally accessible, these principles offer a foundation for engineering structural and electronic order through the deliberate coupling of distinct moiré lattices in multilayer moiré systems.

## Methods

### Sample fabrication

Graphene and hexagonal boron nitride (h-BN) flakes were prepared on $SiO_2$/Si substrates via mechanical exfoliation. Van der Waals (vdW) heterostructures were fabricated using a dry transfer method integrated with a gold-assisted flipping technique[43]. For the tBG/h-BN hetero-trilayers, a polycarbonate (PC)/polydimethylsiloxane (PDMS) stamp was employed to sequentially pick up a bulk h-BN flake followed by two graphene layers. To control the twist angle, twisted bilayer graphene (tBG) was produced by bisecting a single graphene flake using AFM-based cutting[44]. The crystallographic orientations of both bulk h-BN and graphene were identified using well-defined straight edges to ensure reproducible twist-angle control (Fig. S1).

Two distinct twist configurations were prepared:

- Helical configuration: The first graphene layer was rotated by approximately 1° relative to the bulk h-BN, and the second graphene layer was stacked with a rotation in the same direction.
- Alternate configuration: The first graphene layer was rotated by approximately 1° relative to the bulk h-BN, while the second layer was stacked with a rotation in the opposite direction.

The assembled stack was transferred onto a PDMS sheet pre-coated with Au (12 nm)/Ti (3 nm). After removing the PC polymer in N-methyl-2-pyrrolidone (NMP) and isopropyl alcohol (IPA), the heterostructure was flipped and transferred onto a $SiO_2$/Si substrate pre-patterned with metal electrodes. For specific devices, a monolayer or bilayer h-BN was added on top of the tBG without intentional alignment prior to flipping.

### Conductive AFM measurements

Conductive AFM (C-AFM) measurements were performed using a Park Systems NX10 atomic force microscope with $PtIr_5$-coated PPP-EFM cantilevers. Current maps were acquired in contact mode by applying a bias voltage between the tip and the sample. Bias voltage, contact force, and scan rate were optimized for each sample to resolve moiré superlattices (see Table S3 in the SI for specific parameters). We verified that the moiré-moiré reconstruction and stacking registry remained identical across both exposed and h-BN-capped tBG surfaces.

### Data Availability

The data that support the findings of this study are available in the paper and Supplementary Information. Any other data relevant to this paper are available from the corresponding authors on request.

ASSOCIATED CONTENT

**Supporting Information**.

Supplementary Sections I-IX including Supplementary Figs. 1-8.

AUTHOR INFORMATION

**Corresponding Authors**

Correspondence to Y. Seo, N. Nakatsuji, M. Koshino, or T. Machida.

**Author Contributions**

Y.S., J.K. and N.H. performed the C-AFM measurements. Y.S., J.K. and N.H. fabricated the samples. K.W. and T.T. synthesized the h-BN crystals. N.N., T.K. and M.K. carried out the continuum-model simulations. N.N., T.K. and M.K. performed the band-structure calculations. Y.S., N.N. and M.K. analyzed the data. Y.S., N.N., M.K. and T.M. wrote the manuscript with input from all authors. Y.S., N.N., M.K. and T.M. conceived and supervised the project. All authors discussed the results and commented on the manuscript.

**Notes**

The authors declare no competing interests.

ACKNOWLEDGMENTS

This work was supported by JST-CREST, and JST-PRESTO (Grant Numbers JPMJCR20B4, JPMJCR24A5, JPMJPR25H4); JSPS KAKENHI (Grant Numbers JP21H05232, JP21H05233, JP21H05234, JP21H05236, JP23H02052, JP23K26745, JP24K16992, JP23H04862, JP24K21195, and JP25H00602; World Premier International Research Center Initiative (WPI), MEXT, Japan; Joint research funding from Tokyo Electron Limited; Murata Science and Education Foundation. N.N. acknowledges the support from the JSPS Overseas Research Fellowship.

# Supplementary Information:

Twist-configured moiré–moiré reconstruction governs diverse commensurate double-moiré phases in twisted bilayer graphene on h-BN

*Yuta Seo*[1*], *Naoto Nakatsuji*[2*], *Jimpei Kawase*[1], *Naoto Hishida*[1], *Kenji Watanabe*[3], *Takashi Taniguchi*[4], *Takuto Kawakami*[5], *Mikito Koshino*[6] *and Tomoki Machida*[1]

*These authors contributed equally to this work.

[1] Institute of Industrial Science, The University of Tokyo, 4-6-1 Komaba, Meguro, Tokyo 153-8505, Japan

[2] Department of Physics and Astronomy, Stony Brook University, Stony Brook, New York 11794, USA

[3] Research Center for Electronic and Optical Materials, National Institute for Materials Science, 1-1 Namiki, Tsukuba 305-0044, Japan

[4] Research Center for Materials Nanoarchitectonics, National Institute for Materials Science, 1-1 Namiki, Tsukuba 305-0044, Japan

[5] Center for Integrated Science and Humanities, Fukushima Medical University, Fukushima 960-1295, Japan

[6] Institute for Solid State Physics, The University of Tokyo, Kashiwa 277-8581, Japan

## Contents

# I. Sample fabrication

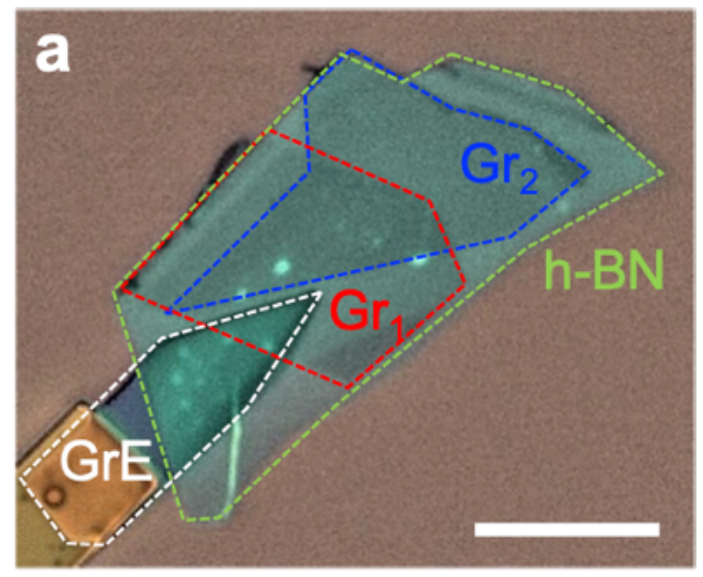

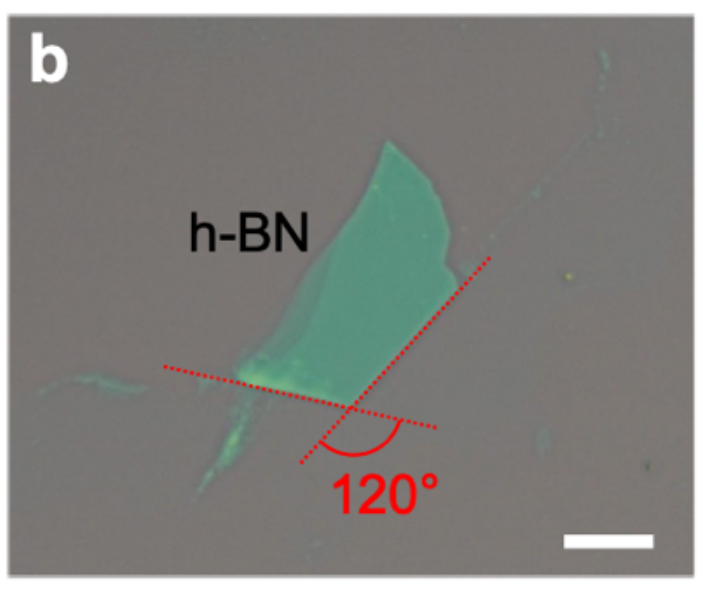

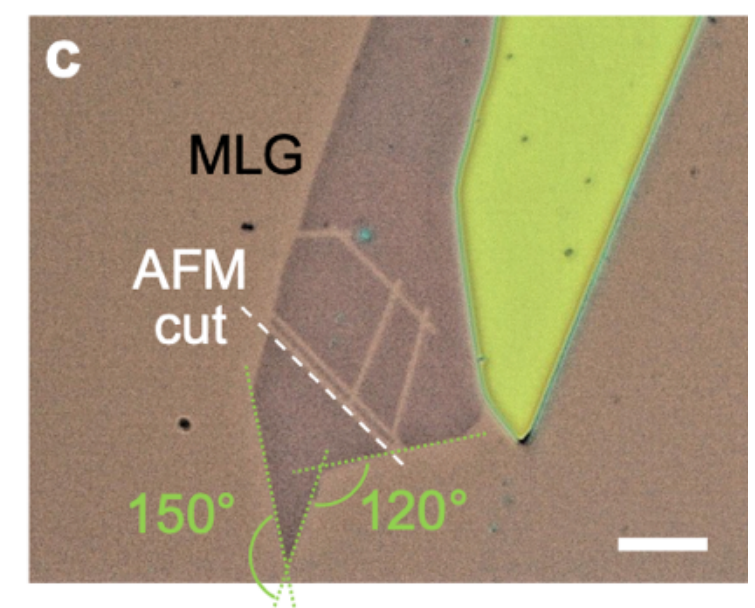

**Figure S1. Sample fabrication. a,** Optical microscope image of the helical tBG/h-BN double moiré sample 1. **b,c,** Optical microscope images of the h-BN flake **(b)** and MLG flake **(c)**. Scale bars correspond to 10 μm.

# II. Theoretical model

## A. Geometry

We introduce the lattice geometry of twisted bilayer graphene aligned h-BN. To include the thickness effect of h-BN of the experiment, we consider the non-twisted trilayer h-BN in this paper. The graphene layers are denoted as layers 1 and 2, while the h-BN layers are denoted as layer 3, 4, and 5 as shown in Figure S2. For simple notation, in this section, we use $\theta^{12}$ and $\theta^{23}$ as the twist angles $\theta_{\mathrm{tBG}}$ and $\theta_{\mathrm{Gr/BN}}$ respectively.

The reciprocal vectors of moiré patterns between layer 1 and 2, and between layer 2 and 3 are obtained from $\boldsymbol{G}_j^{\ell\ell'} = \boldsymbol{b}_j^{(\ell)} - \boldsymbol{b}_j^{(\ell')}$ for $(\ell,\ell') = (1,\ 2)$ and $(2,\ 3)$, where $\boldsymbol{b}_j^{(\ell)} = R(\theta^\ell)(1+\epsilon_\ell)^{-1}\boldsymbol{b}_j$ is the atomic reciprocal lattice vector of layer $\ell$, and $\boldsymbol{b}_1 = (2\pi/a)\ (1, 1/\sqrt{3})$ and $\boldsymbol{b}_2 = (2\pi/a)\ (0,1)$ are the reciprocal lattice vectors of graphene without rotation. $R(\theta)$ is a two-dimensional rotation matrix, and $\theta^\ell$ is the twist angle of layer $\ell$ specified by $\theta^{(1)} = -\theta^{12}, \theta^{(2)} = 0$ and $\theta^{(3)} = \theta^{23}$. $\epsilon_\ell = (a_{BN} - a)/a$ for $\ell = 3{,}4$, and 5, otherwise 0, where $a = 2.46$ Å and $a_{BN} = 2.504$ Å are the atomic lattice constant of graphene and h-BN respectively. The moiré lattice vectors in the real space are given by $\boldsymbol{G}_i^{\ell\ell'} \cdot \boldsymbol{L}_j^{\ell\ell'} = 2\pi\delta_{ij}$, and corresponding moiré length is written as $\left|\boldsymbol{L}_j^{\ell\ell'}\right| = 4\pi/(\sqrt{3}\left|\boldsymbol{G}_j^{\ell\ell'}\right|)$.

Generally, above two moiré lattices are incommensurate. However, we can define the approximate moiré-of-moiré (MoM) period when there is a certain points which two moiré patterns are almost aligned each other. Such a situation is expressed as

$$\begin{pmatrix} \boldsymbol{L}_1 \\ \boldsymbol{L}_2 \end{pmatrix} = \begin{pmatrix} n_1^{12} & m_1^{12} \\ n_2^{12} & m_2^{12} \end{pmatrix} \begin{pmatrix} \boldsymbol{L}_1^{12} \\ \boldsymbol{L}_2^{12} \end{pmatrix} = \begin{pmatrix} n_1^{23} & m_1^{23} \\ n_2^{23} & m_2^{23} \end{pmatrix} \begin{pmatrix} \boldsymbol{L}_1^{23} \\ \boldsymbol{L}_2^{23} \end{pmatrix} + \begin{pmatrix} \Delta\boldsymbol{L}_1 \\ \Delta\boldsymbol{L}_2 \end{pmatrix} \tag{1}$$

for integers $n_j^{\ell\ell'}$ and $m_j^{\ell\ell'}$, and the difference $\Delta\boldsymbol{L}_j$. From $\boldsymbol{L}_2 = R(60°)\boldsymbol{L}_1$, the integers for $\boldsymbol{L}_2$ is written by $n_2^{\ell\ell'} = -n_1^{\ell\ell'}$ and $m_2^{\ell\ell'} = n_1^{\ell\ell'} + m_1^{\ell\ell'}$, and $\Delta\boldsymbol{L}_2 = R(60°)\Delta\boldsymbol{L}_1$. For smaller $\left|\Delta\boldsymbol{L}_j\right|$ than the moiré periods, the system can be approximated by neglecting $|\Delta \boldsymbol{L}_j|$. Specifically, in this paper, it is obtained by slightly rotating and expanding/shrinking of the h-BN layer so that $|\Delta \boldsymbol{L}_j|$ vanishes. Such exact commensurate lattice gives the $|\boldsymbol{L}_1^{12}|:|\boldsymbol{L}_1^{23}| = \sqrt{N_1^{23}}:\sqrt{N_1^{12}}$ commensurate structure, where $N_j^{\ell\ell'} = \left(n_j^{\ell\ell'}\right)^2 + n_j^{\ell\ell'} m_j^{\ell\ell'} + \left(m_j^{\ell\ell'}\right)^2$. The corresponding reciprocal lattice vectors for the exact commensurate lattice are given by

$$\begin{pmatrix} \boldsymbol{G}_1 \\ \boldsymbol{G}_2 \end{pmatrix} = \begin{pmatrix} n_1^{12} & m_1^{12} \\ n_2^{12} & m_2^{12} \end{pmatrix}^{-1} \begin{pmatrix} \boldsymbol{G}_1^{12} \\ \boldsymbol{G}_2^{12} \end{pmatrix} = \begin{pmatrix} n_1^{23} & m_1^{23} \\ n_2^{23} & m_2^{23} \end{pmatrix}^{-1} \begin{pmatrix} \boldsymbol{G}_1^{23} \\ \boldsymbol{G}_2^{23} \end{pmatrix} \tag{2}$$

For the $C_{3z}$-symmetric commensurate structure considered in the main text, we estimate the twist angles by matching the moiré periods and its commensurate configurations observed in AFM images. We select the angles that best reproduce the experimental patterns. The estimated twist angles are listed on Table S1.

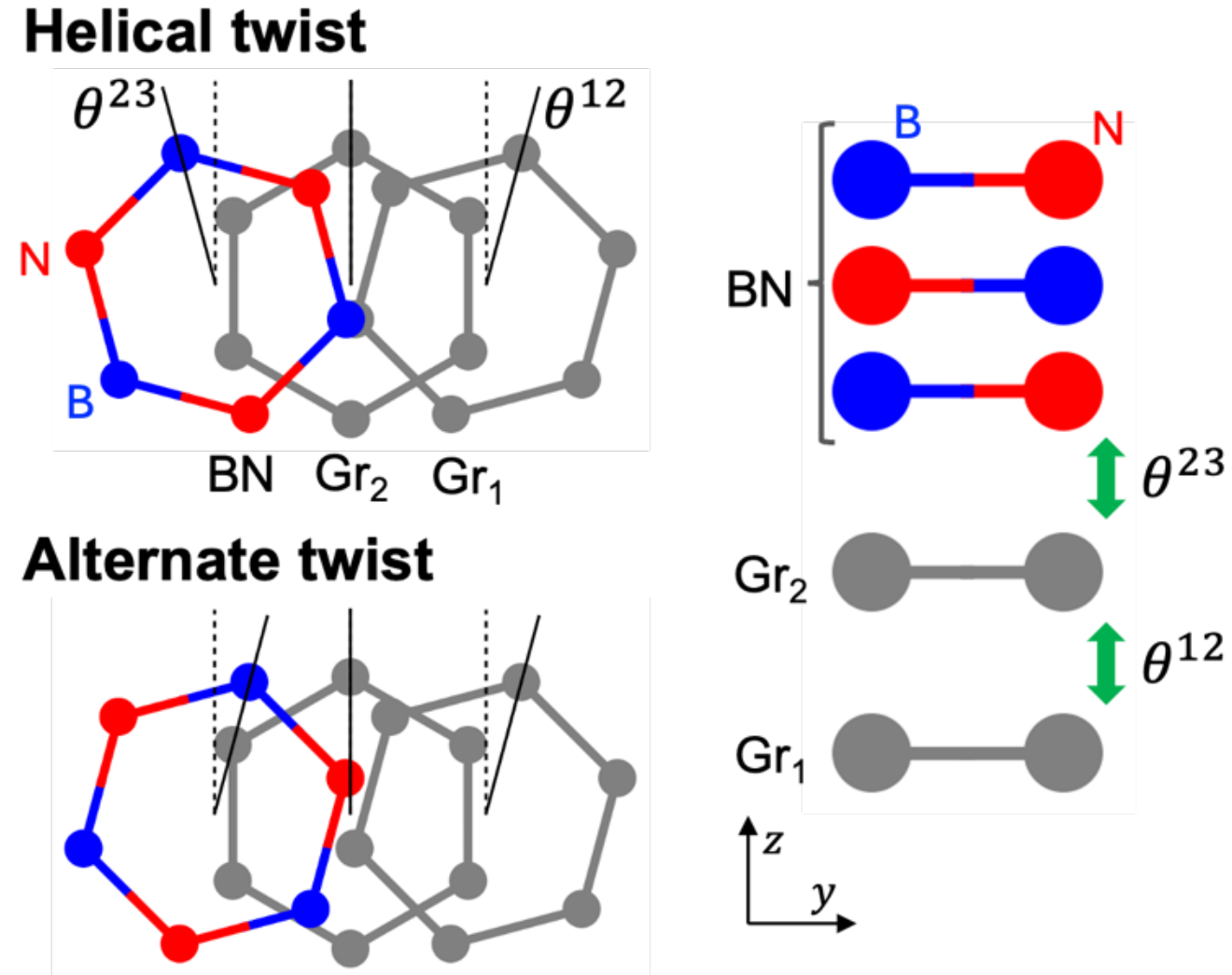

**Figure S2.** Schematic illustration of tBG on three layer h-BN. The left panels show top views of the helical (top) and alternate (bottom) twist configurations, where gray, red, and blue dots represent carbon, nitrogen, and boron atoms, respectively. The right panel shows a line cut in the *y*-*z* plane.

**Table S1.** Correspondence between figure indices, domain types defined by the moiré length ratio $|\boldsymbol{L}_1^{12}|:|\boldsymbol{L}_1^{23}|$, the estimated twist angles $(\theta^{12}, \theta^{23})$, and the integer sets $(n_1^{12}, m_1^{12}, n_1^{23}, m_1^{23})$ used in Eq. 1.

| Figure | Domain type | $(\theta^{12}, \theta^{23})$ [∘] | $(n_1^{12}, m_1^{12}, n_1^{23}, m_1^{23})$ |
|---|---|---|---|
| 3a | 1 : 1 | (1.16, 0.60) | (1, 0, 1, −1) |
| 3b | 1 : 1 | (1.01, 0.52) | (1, 0, 1, −1) |
| 1a, 3c | 2 : 1 | (0.60, 0.62) | (1, 0, 2, −2) |
| 3d | 3 : 1 | (0.42, 0.64) | (1, 0, 3, −3) |
| 1g, 3f | $\sqrt{7}$ : 1 | (0.39, −0.97) | (1, 0, −3, 1) |
| 3i | 2 : $\sqrt{3}$ | (0.87, $-7.6 \times 10^{-3}$) | (1, 1, 2, −2) |

**Table S2.** Correspondence between figure indices, domain types for strained configurations, the estimated twist angles $(\theta^{12}, \theta^{23})$, and integers $\left(n_j^{12}, m_j^{12}, n_j^{23}, m_j^{23}\right)$ used Eq. 1. The domain types are defined by the moiré length ratios $|\boldsymbol{L}_1^{12}|:|\boldsymbol{L}_1^{23}|;|\boldsymbol{L}_2^{12}|:|\boldsymbol{L}_2^{23}|$.

| Figure | Domain type | $(\theta^{12}, \theta^{23})$ [$\circ$] | $(n_1^{12}, m_1^{12}, n_1^{23}, m_1^{23})$ | $(n_2^{12}, m_2^{12}, n_2^{23}, m_2^{23})$ |
|---|---|---|---|---|
| 3e | $2:1; 3:1$ | (0.53, 0.75) | (1, 0, 3, −2) | (0, 1, 3, 0) |
| 3g | $\sqrt{3}:1; 2:1$ | (0.54, −0.25) | (1, 0, 1, −3) | (0, 1, 2, −2) |
| 3h | $\sqrt{7}:1; \sqrt{13}:1$ | (0.30, −0.41) | (1, 0, 2, −3) | (0, 1, 4, −1) |
| 3j | $\sqrt{7}:2; \sqrt{3}:1$ | (0.73, $-7.3 \times 10^{-3}$) | (2, 0, 2, −3) | (0, 1, 2, −1) |

We also consider strained cases in the main text. We define the exact commensurate moiré of-moire unit cell based on the MoM patterns observed in the experimental AFM images. From the AFM images, we find that the graphene/h-BN moiré pattern exhibits a larger distortion than the tBG moiré pattern. Based on this observation, we assume that layers 2 and 3, which form the graphene/h-BN moiré, experience homo-strain, while layers 1 and 2, which form tBG, experience hetero-strain.

Under this assumption, we write the strained reciprocal lattice vector of layer $\ell$ as

$$\boldsymbol{b}_j^{\ell} = (I_2 + s_\ell \mathcal{E}) R\left(\theta^{(\ell)}\right) \boldsymbol{b}_j \tag{3}$$

where

$$\mathcal{E} = \begin{pmatrix} \epsilon_+ + \epsilon_- & \epsilon_{xy} - \Omega \\ \epsilon_{xy} + \Omega & \epsilon_+ - \epsilon_- \end{pmatrix} \tag{4}$$

is the strain tensor, with $s_{\ell=1} = -1$ and $s_{\ell=2,3} = 1$. Here, $\epsilon_\pm$ and $\epsilon_{xy}$ describe the strain components, while $\Omega$ represents the rotational deformation.

To estimate the twist angles, we first determine the integers $n_j^{\ell\ell'}, m_j^{\ell\ell'}$ from the AFM images. For assumed twist angles $\theta^{12}$ and $\theta^{23}$, we then define the moiré reciprocal lattice vectors $\boldsymbol{G}_j^{\ell\ell'} = \boldsymbol{b}_j^{(\ell)} - \boldsymbol{b}_j^{(\ell')}$ as functions of the strain components by using Eq. (3). Substituting these into Eq. (1) while neglecting $\Delta\boldsymbol{L}_j$, we solve for $\epsilon_\pm, \epsilon_{xy}$, and $\Omega$. This procedure is repeated for several assumed twist angles, and we select those that yield lattice periods closest to the AFM patterns. The estimated twist angles are summarized in Table S2.

### B. Commensurate Domain Phase Diagram

Since large MoM pattern is significantly modulated by the small distortion of the moiré lattice, it is expected that MoM materials form domains when the generalized MoM unit cell becomes sufficiently large. Such large MoM pattern appears when two moiré patterns are approximately commensurate i.e. $\Delta L_j$ in Eq.

(1) is small enough. In Fig. 3l of main text, we highlight the regions where the generalized MoM lattice satisfies $\frac{|\Delta \boldsymbol{L}_j|}{\max\left(\left|\boldsymbol{L}_j^{12}\right|,\left|\boldsymbol{L}_j^{23}\right|\right)} < 0.1$. This criterion is motivated by our previous study of twisted trilayer graphene[1], in which commensurate domains were observed even for relatively small MoM periods (e.g., $\frac{|\Delta \boldsymbol{L}|}{\max(|\boldsymbol{L}^{12}|,|\boldsymbol{L}^{23}|)} \sim 0.20$), and thus provides a conservative lower bound for domain formation.

### C. Continuum Method for optimized lattice structure

To simulate the optimized lattice structure, we apply the elastical continuum method[1,2]. This model we numerically find the displacement vector $u^{\ell}$ minimizing the total energy $U = U_E + \sum_{\ell=1}^{4} U_B^{\ell,\ell+1}$. The elastic energy $U_E$ is written as[3,4]

$$U_E = \sum_{\ell=1}^{5} \frac{1}{2} \int d^2\boldsymbol{r} \left[ \left(\mu^{(\ell)} + \lambda^{(\ell)}\right)\left(u_{xx}^{(\ell)} + u_{yy}^{(\ell)}\right)^2 + \mu^{(\ell)} \left\{ \left(u_{xx}^{(\ell)} - u_{yy}^{(\ell)}\right)^2 + 4\left(u_{xy}^{(\ell)}\right)^2 \right\} \right], \tag{6}$$

Here $u_{\mu\nu} = \left(\partial_\mu u_\nu + \partial_\nu u_\mu\right)/2$ is the strain tensor. We use the Lamé parameters $\lambda^{(\ell=1,2)} = 325$ eV/nm$^2$ and $\mu^{(\ell=1,2)} = 957$ eV/nm$^2$ for graphene[5,6], and $\lambda^{(\ell=1,2)} = 350$ eV/nm$^2$ and $\mu^{(\ell=1,2)} = 780$ eV/nm$^2$ for h-BN[6,7]. Experimentally, we find that the graphene/h-BN moiré pattern is relatively rigid. Therefore, when not otherwise specified, we take enhanced effective Lamé parameters $\tilde{\lambda}^{(\ell)} = 2.4\,\lambda^{(\ell)}$ and $\tilde{\mu}^{(\ell)} = 2.4\,\mu^{(\ell)}$ for $\ell = 2,\ 3,\ 4,$ and 5. The interlayer moiré binding energy of adjacent layers $(\ell, \ell') = (1,2)$ and $(2,3)$ is given by[8]

$$U_B^{\ell\ell'} = \int d^2\boldsymbol{r} \sum_{j=1}^{3} 2U_0^{\ell\ell'} \cos\left[G_j^{\ell\ell'} \cdot r + b_j \cdot \left(u^{(\ell')} - u^{(\ell)}\right) + \phi_0^{\ell\ell'}\right], \tag{7}$$

where $\boldsymbol{b}_3 = -\boldsymbol{b}_1 - \boldsymbol{b}_2$ and $\boldsymbol{G}_3^{\ell\ell\prime} = -\boldsymbol{G}_1^{\ell\ell'} - \boldsymbol{G}_2^{\ell\ell'}$. The interlayer binding energy of h-BN layers $(\ell, \ell') = (3,4)$ and $(4,5)$ is given by

$$U_B^{\ell\ell'} = \int d^2\boldsymbol{r} \sum_{j=1}^{3} 2U_{\mathrm{BN}} \cos\left[b_j \cdot \left(\boldsymbol{u}^{(\ell')} - \boldsymbol{u}^{(\ell)}\right) + \phi_{BN}^{\ell\ell'}\right], \tag{8}$$

where $\phi_{BN}^{34} = -\phi_{BN}^{45}$ due to 2H stacking of h-BN. For parameters, we set $U_0^{12} = 0.160$ eV/nm$^2$, $U_0^{23} = 0.202$ eV/nm$^2$, $\phi_0^{12} = 0$, and $\phi_0^{23} = -956$ for tBG/h-BN[9–12], and $\phi_{BN}^{34} = -0.888$ and $U_{BN} = -0.072$ eV/nm$^2$ for nontwist h-BN. To obtain the anti-parallel h-BN binding energy, we used the local stacking energy $V_{AA'} = 0$ eV/nm$^2$, $V_{AB'} = 0.699$ eV/nm$^2$, and $V_{BA'} = 0.119$ eV/nm$^2$ [13].

### D. Continuum Hamiltonian

We introduce the continuum Hamiltonian for tBG/h-BN by extending the model for trilayer graphene[1]. From here, since only graphene electron is dominant around Fermi energy, we only consider the top most h-BN (layer 3) and ignore the layer 4 and 5. The effective Hamiltonian for valley $\xi$ is written as

$$H^{(\xi)} = \begin{pmatrix} H_1(\boldsymbol{k}) & U_{21}^\dagger & \\ U_{21} & H_2(\boldsymbol{k}) & U_{32}^\dagger \\ & U_{32} & H_3(\boldsymbol{k}) \end{pmatrix}. \quad (9)$$

The basis is taken as $\left(\psi_A^{(1)}, \psi_B^{(1)}, \psi_A^{(2)}, \psi_B^{(2)}, \psi_A^{(3)}, \psi_B^{(3)}\right)$, where $\psi_X^{(\ell)}$ represents the envelope function of sublattice $X = A, B$ on layer $l = 1, 2, 3$. The $H_\ell(k)$ is the 2 × 2 Hamiltonian of monolayer graphene and h-BN. $U_{\ell\ell'}$ is the interlayer coupling matrix, in the presence of the lattice distortion. The $H_\ell(k)$ is given by

$$H_\ell(\boldsymbol{k}) = -hv\left[R\left(-\theta^{(\ell)}\right)\left(\boldsymbol{k} - \boldsymbol{K}_\xi^{(\ell)} + \tfrac{e}{\hbar}\boldsymbol{A}^{(\ell)}\right)\right] \cdot \boldsymbol{\sigma} + \delta_{3,\ell}\begin{pmatrix} V_N & \\ & V_B \end{pmatrix}, \quad (10)$$

where $v$ is the graphene's band velocity, $\boldsymbol{\sigma} = (\xi\sigma_x, \sigma_y)$ and $\sigma_x, \sigma_y$ are the Pauli matrix. $K^{(\ell)} = -\xi\left(2\, b_1^{(\ell)} + b_1^{(\ell)}\right)/3$ is the $\xi$-valley Dirac point of layer $\ell$.

We take $\hbar v/a = 2.14$ eV [14,15]. The $A^{(l)}$ is the strain-induced vector potential that is given by[3,16,17]

$$\boldsymbol{A}^{(\ell)} = \xi\frac{3}{4}\frac{\beta\gamma_0}{ev}\begin{pmatrix} u_{xx}^{(\ell)} - u_{yy}^{(\ell)} \\ -2u_{xy}^{(\ell)} \end{pmatrix}, \quad (11)$$

where $\gamma_0 = 2.7$ eV is the nearest neighbor transfer energy of intrinsic graphene and $\beta \approx 3.14$ [15].

The interlayer coupling matrix $U_{21}$ and $U_{32}$ are given by

$$U_{\ell'\ell} = \sum_{j=1}^{3} U_j^{\ell'\ell}\, e^{i\xi\delta\boldsymbol{k}_j^{\ell\ell'}\cdot\boldsymbol{r} + i\,\boldsymbol{Q}_j\cdot\left(\boldsymbol{u}^{(\ell')} - \boldsymbol{u}^{(\ell)}\right)}, \quad (12)$$

where we defined

$$\delta\boldsymbol{k}_1^{\ell\ell'} = 0\,,\ \delta\boldsymbol{k}_2^{\ell\ell'} = \xi\boldsymbol{G}_1^{\ell\ell'},\ \delta\boldsymbol{k}_3^{\ell\ell'} = \xi\left(\boldsymbol{G}_1^{\ell\ell'} + \boldsymbol{G}_2^{\ell\ell'}\right),$$

$$\boldsymbol{Q}_1 = \boldsymbol{K}_\xi\,,\ \boldsymbol{Q}_2 = \boldsymbol{K}_\xi + \xi\boldsymbol{b}_1,\ \boldsymbol{Q}_3 = \boldsymbol{K}_\xi + \xi(\boldsymbol{b}_1 + \boldsymbol{b}_2), \quad (13)$$

and

$$U_j^{\ell\ell'} = u_{\ell\ell'}I_2 + u'_{\ell\ell'}\left(\sigma_x\cos\phi_j + \sigma_y\sin\phi_j\right), \quad (14)$$

where $\phi_j = \xi 2\pi(j-1)/3$. The parameters $u_{\ell\ell'}$ and $u_{\ell\ell'}$are interlayer coupling strength of AA and AB/BA stacking between layer $\ell$ and $\ell'$. $K_\xi$ is the Dirac point without the rotation. We take $u_{12} = 79.7$ meV and $u'_{12} = 95.7$ meV for tBG[18], and $u_{23} = u'_{23} = 152$ meV for Gr/h-BN[19].

## III. Moiré-of-moiré scale calculation

In the main text, we theoretically focus on commensurate configurations realized at specific twist angles and strain conditions. In a realistic trilayer system, however, the two constituent moiré patterns generally have a slight mismatch, leading to the formation of a moiré-of-moiré (MoM) structure. Here, we perform MoM-scale calculations and show that the commensurate configurations can locally emerge through lattice relaxation.

Figure S3 shows the lattice structures of tBG/h-BN for both the helical and alternate twist configurations, where the blue and red patterns represent the triangular moiré lattice of tBG and the hexagonal moiré lattice arising from graphene/h-BN, respectively. The blue dots indicate AA stacking regions of tBG, while the interior of the red hexagons corresponds to AB stacking regions of graphene/h-BN. We choose twist angles close to the $2:1$ and $\sqrt{7}:1$ commensurate conditions for the helical and alternating cases, respectively.

By comparing the non-relaxed [Figure S3(a)] and relaxed [Figure S3(b)] structures in the helical case, we find that lattice relaxation favors the alignment between the AA region of tBG and the AB′ region of graphene/h-BN, thereby spontaneously forming domains of the local $2:1$ commensurate configuration. In the alternate case, lattice relaxation similarly stabilizes the moiré stacking between the AA region of tBG and the AA′/BA′ regions of graphene/h-BN, resulting in local structures closely resembling the $\sqrt{7}:1$ commensurate configuration [see Figure S3(c,d)].

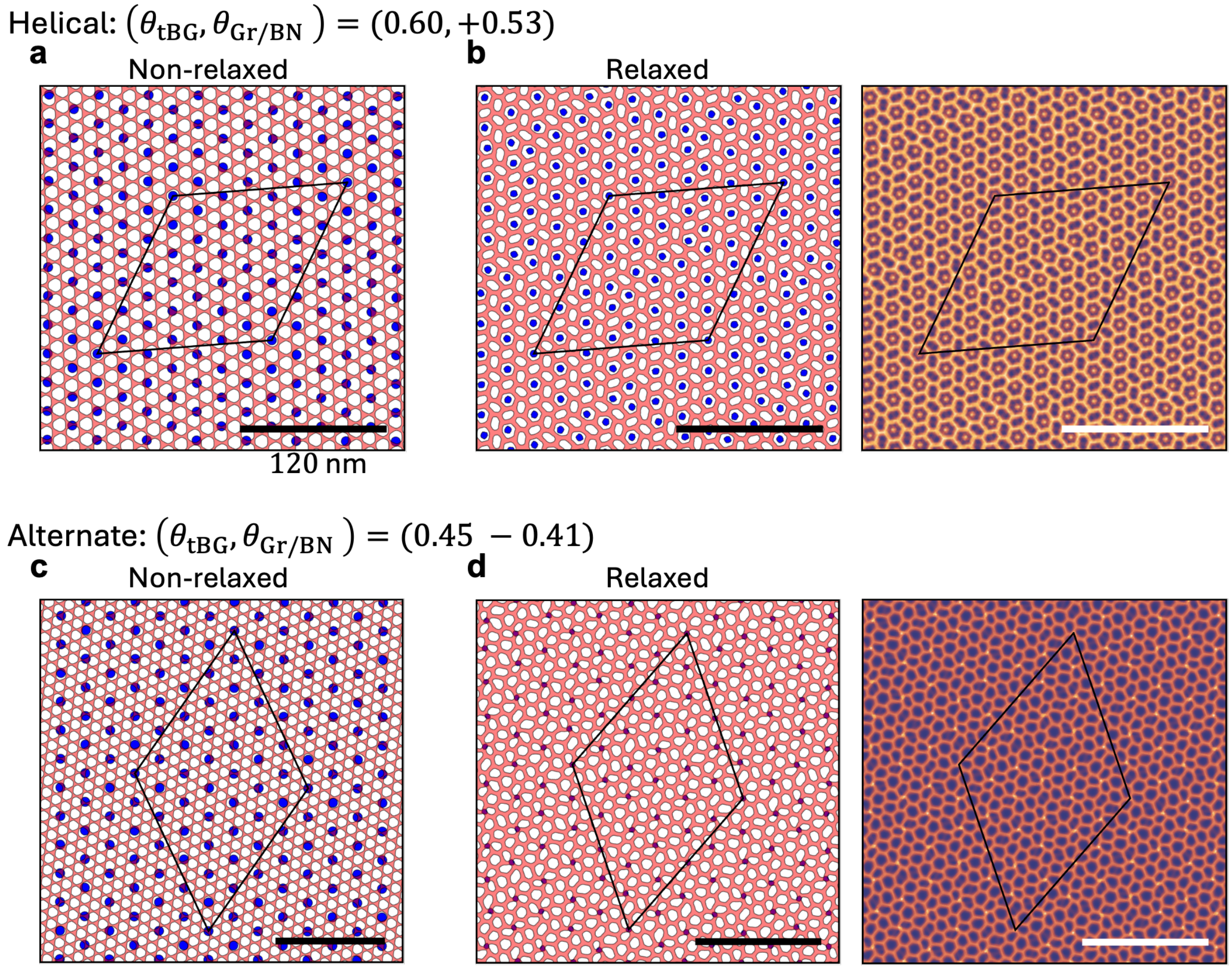

**Figure S3. Simulated lattice structures of tBG/h-BN at the MoM scale. a**, Non-relaxed lattice structure of tBG/h-BN with $(\theta_{\mathrm{tBG}}, \theta_{\mathrm{Gr/BN}}) = (0.60°, 0.53°)$, close to the $2:1$ configuration, where blue and red patterns represent the moiré lattices of tBG and graphene/h-BN, respectively. **b**, Relaxed structure, showing (left) the schematic configuration and (right) the real-space distribution of the binding energy. **c–d**, Analogous plots for the alternating twist case with $(\theta_{\mathrm{tBG}}, \theta_{\mathrm{Gr/BN}}) = (0.45°, -0.41°)$, close to the $\sqrt{7}:1$ configuration. In all panels, the black rhombus denotes the MoM unit cell, and the scale bar corresponds to 120 nm. The calculations are performed using the bare Lamé parameters.

## IV. Strained commensurate configuration

As mentioned in the main text, we observe strained commensurate configurations [see Fig. 3 in the main text]. These strained configurations can be understood as arising from $C_{3z}$-symmetric configurations neighboring each other in the phase diagram shown in Fig. 3k of the main text. Fig. S4a shows an enlarged view of Fig. 3e for the helical twist case. Fig. S4c shows a schematic of the strained configuration, where the filled green and orange dots represent AA stacking of tBG and AA′/BA′ stacking of graphene/h-BN, respectively. The open circles indicate AA stacking of tBG in the $C_{3z}$-symmetric $L_{\mathrm{tBG}}:L_{\mathrm{Gr/BN}} = 3:1$ configuration.

For helical-twist tBG/h-BN, AA stacking of tBG tends to align with AB′ stacking (the center of the hexagon) of graphene/h-BN due to local rotational matching [see main text]. When a global strain is present, this alignment leads the system to favor a strained configuration in which AA-stacked regions of tBG shift to the next AB′ sites of graphene/h-BN, as shown in Fig. S4c. As shown in Fig. S4d, the $C_{3z}$-symmetric $2:1$ configuration serves as another parent structure. The strained configurations can therefore be understood as originating from nearby $C_{3z}$-symmetric points in the domain phase diagram. In particular, the strained configurations in Fig. 3g and 3h arise from the $\sqrt{3}:1$ and $2:1$ configurations, and from the $\sqrt{7}:1$ and $\sqrt{13}:1$ configurations, respectively. This behavior is unique to tBG/h-BN and is distinct from tBG, which exhibits a continuously deformed AA-stacked configuration under strain.

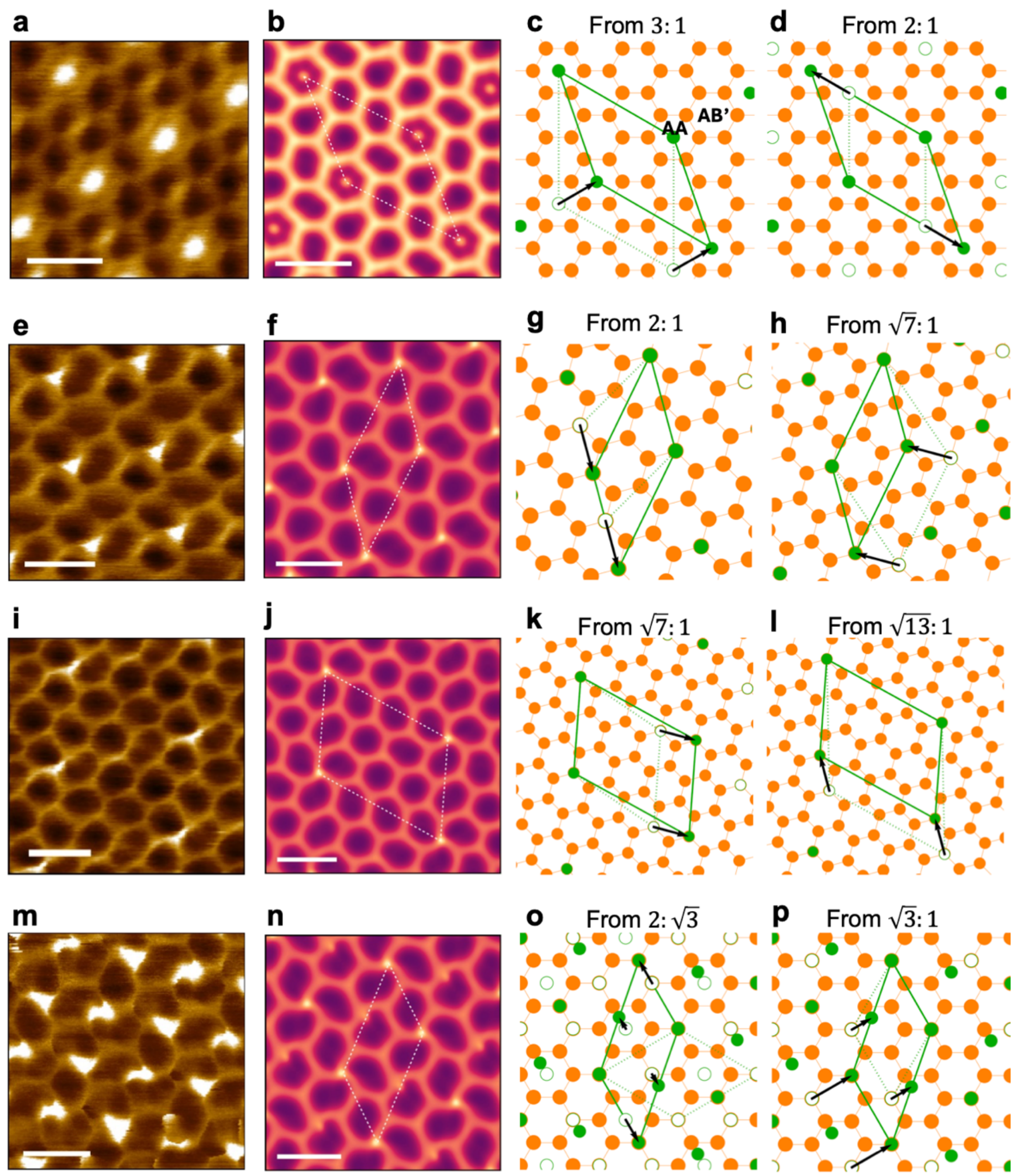

**Figure S4. Various strained configurations originating from the $C_{3z}$-symmetric state. a**, Experimental C-AFM image corresponding to an enlarged view of Fig. 3e in the main text. **b**, Simulated structure obtained from the continuum model. **c–d**, Schematic illustrations of the configuration transition from the $C_{3z}$-symmetric states with $L_{\mathrm{tBG}}: L_{\mathrm{Gr/BN}} = 3:1$ (c) and 2: 1 (d), respectively. Green and orange filled dots represent AA stacking in tBG and AA′ /BA′ stacking in graphene/h-BN for the strained configurations. Open circles indicate the corresponding $C_{3z}$-symmetric configurations for the 3: 1 and 2: 1 cases, respectively. Solid and dashed lines denote the unit cells of the strained and $C_3z$-symmetric configurations. **e–h**, Similar panels as in **a–d** for the strained configuration in Fig. 3g. **i–l**, Similar panels for Fig. 3h. **m–p**, Similar panels for Fig. 3j.

## V. Formation conditions for commensurate double-moiré domains

In double-moiré trilayer systems, the two moiré lattices can spontaneously form commensurate domains even when the twist angles are not exactly commensurate. The formation of such domains via local rotational matching mechanism requires two conditions: (1) both moiré patterns undergo local rotational relaxation, and (2) the two moiré lattices are nearly commensurate i.e. having small $|\Delta L|$ in Eq. (1), as reflected in the emergence of a large moiré-of-moiré period.

The first condition follows from the fact that the local rotational matching mechanism is driven by rotational components of lattice relaxation in both moiré patterns. The second condition arises because a larger moiré-of-moiré period can be more readily modulated by small lattice distortions, analogous to the fact that lattice relaxation becomes increasingly significant at smaller twist angles in bilayer moiré systems. In tBG/h-BN, these conditions impose restrictive constraints on the allowed twist angles. In particular, the first condition becomes difficult to satisfy near $\theta \sim 0°$ for hetero-bilayer moiré, including Gr/h-BN, where relaxation is dominated by local expansion and contraction rather than rotation[6,12]. As a result, in tBG/h-BN, the formation of large commensurate domains is hindered near $\theta_{\mathrm{Gr/BN}} \sim 0°$, as illustrated in Fig. 3i, which exhibits more strongly strained structures compared to other commensurate patterns. At larger twist angles, rotational relaxation of bilayer moiré becomes progressively suppressed (typically above $\sim 3$ degrees[12]), which further limits the formation of commensurate domains. The second condition is likewise more restrictive than tTG, Homo double-moiré systems, because the moiré-of-moiré period depends sensitively on the twist angles $\theta_{\mathrm{Gr/BN}}$ when the Gr/h-BN moiré itself rotates with them. Consequently, commensurate domain formation occurs within narrower regions of parameter space than in homo double-moiré systems.

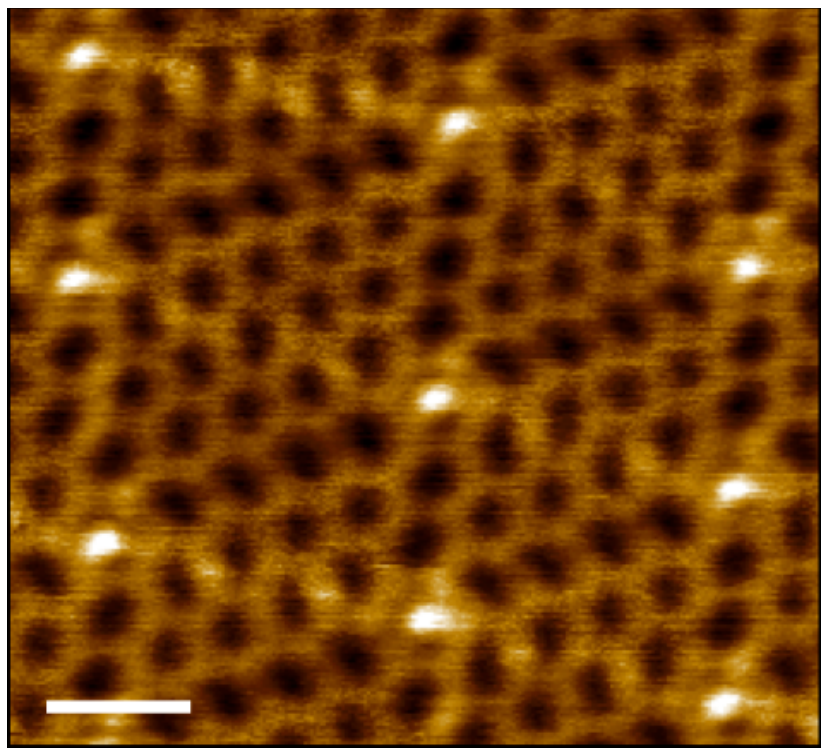

**Figure S5. Double-moiré structure without long-range commensurate order in the helical configuration.** Experimental C-AFM image showing a region where the local moiré–moiré stacking registry is preserved but the period ratio $L_{\mathrm{tBG}}/L_{\mathrm{Gr/BN}}$ varies gradually across the field of view, indicating the absence of long-range commensurate domain formation. Scale bar, 20 nm.

## VI. Domain-boundary morphology and relative moiré-unit density

To complement the large-scale ordering behavior discussed in the main text, we further examine the morphology of domain boundaries in tBG/h-BN double-moiré lattices as a function of the relative moiré-unit density, $N_{\mathrm{tBG}}$ and $N_{\mathrm{Gr/BN}}$, using the data shown in Figure S6. Figure S6c and Figure S6d compare representative helical configurations with different relative densities of tBG AA domains and available Gr/h-BN template sites. When the density of tBG AA domains is smaller than that of the favorable Gr/h-BN domains (Figure S6c; $N_{\mathrm{tBG,AA}} < N_{\mathrm{Gr/BN,AB'}}$), the system forms extended, nearly straight domain boundaries, consistent with an ordered domain morphology. In contrast, when the tBG AA density exceeds the available template capacity (Figure S6d; $N_{\mathrm{tBG,AA}} > N_{\mathrm{Gr/BN,AB'}}$), the boundary network becomes irregular and fragmented, and long straight boundaries are no longer sustained. This contrast suggests that the formation of straight, extended boundaries is favored when the tBG moiré can be accommodated by the underlying Gr/h-BN scaffold without accumulating excess AA domains. When such accommodation is incomplete, we believe the additional AA domains are driven toward the boundaries, leading to local frustration and the emergence of disordered boundary networks. The corresponding schematic illustrations in Figure S6e and Figure S6f summarizes this qualitative picture. In the ordered case (Figure S6e), the uniform Gr/h-BN moiré lattice provides sufficient template sites to lock the tBG AA domains, resulting in straight boundaries separating well-defined domains. In the disordered case (Figure S6f), we believe the mismatch between the number of tBG AA domains and the available template sites forces distortions to concentrate along the boundaries, producing irregular and zigzag-like boundary morphologies. We emphasize that these observations represent a phenomenological correlation between boundary morphology and relative moiré-unit density. While the relative density provides a useful organizing principle for classifying the observed domain structures, a quantitative theoretical description of the density-dependent transition is beyond the scope of the present work.

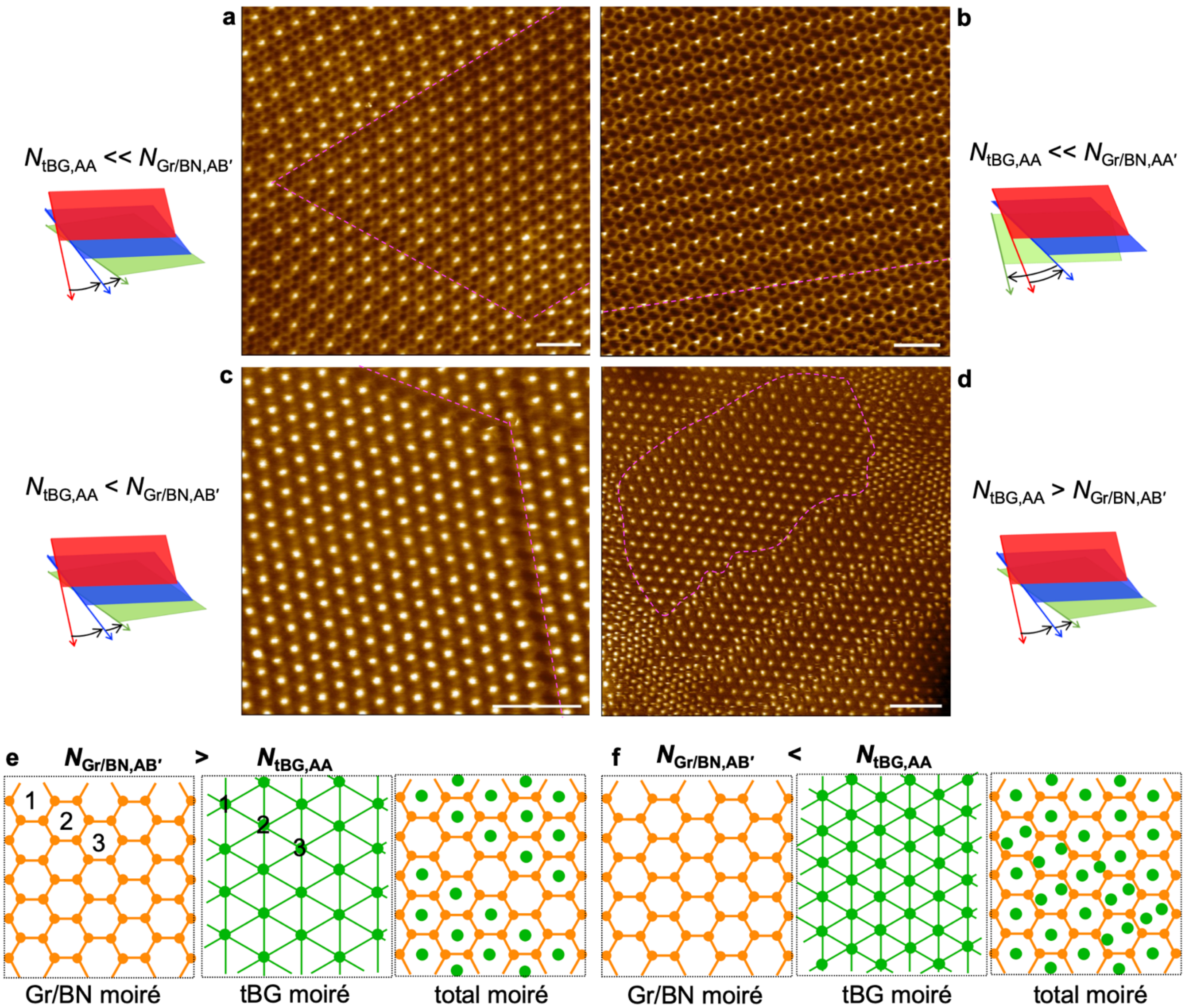

**Figure S6. Long-range domain morphology governed by the relative moiré densities. a-d,** Large-area C-AFM images showing domain morphology for representative helical and alternate configurations with different ratios of $N_{tBG}$ and $N_{Gr/BN}$. **a,** Helical sample with $N_{tBG,AA} << N_{Gr/BN,AB'}$, showing extended linear boundaries. **b**, Alternate sample with $N_{tBG,AA} << N_{Gr/BN,AA'}$, also exhibiting long, straight boundaries. **c**, Helical case with $N_{tBG,AA} < N_{Gr/BN,AB'}$, also exhibiting long, straight boundaries. **d,** Helical case with $N_{tBG,AA} > N_{Gr/BN,AB'}$, displaying irregular and disordered domain boundaries. Scale bars correspond to 50 nm. **e,f,** Schematics illustrating the contrast between ordered ($N_{tBG,AA} < N_{Gr/BN,AB'}$) and disordered ($N_{tBG,AA} > N_{Gr/BN,AB'}$) domain morphologies. These results identify the relative moiré-unit density as the key parameter controlling long-range order.

## VII. Impact of the twist configuration—helical and alternate—on the electronic band structure

The twist arrangement (helical or alternate) has a significant impact on the electronic band structure. Figure S7 compares the electronic band structures for helical and alternating arrangements at the same tBG twist angle $\theta_{\mathrm{tBG}}$. In each case, the left panel shows the optimized lattice structure with the $2:1$ configuration, while the right panel shows the corresponding electronic band structure and density of states (DOS).

Due to the difference in the optimized lattice arrangements, the electronic band structure is strongly modified, including changes in band flatness and the Chern number of the bands. This result indicates that the twist arrangement in tBG/h-BN provides an important tuning knob for controlling both band dispersion and topology.

The origin of these differences can be qualitatively understood from the distinct local stacking environments realized in the two configurations. In particular, the flat-band states in tBG are known to be localized near AA regions, and thus are sensitive to the effective potential induced by the underlying hBN layer. As shown in Figure S7 [also discussed in main text], helical and alternate arrangements favor different relative alignments between the tBG AA regions and the graphene/h-BN moiré pattern, leading to different spatial distributions of the h-BN-induced potential. This difference likely results in distinct modifications of the flat bands, including their bandwidth and topology.

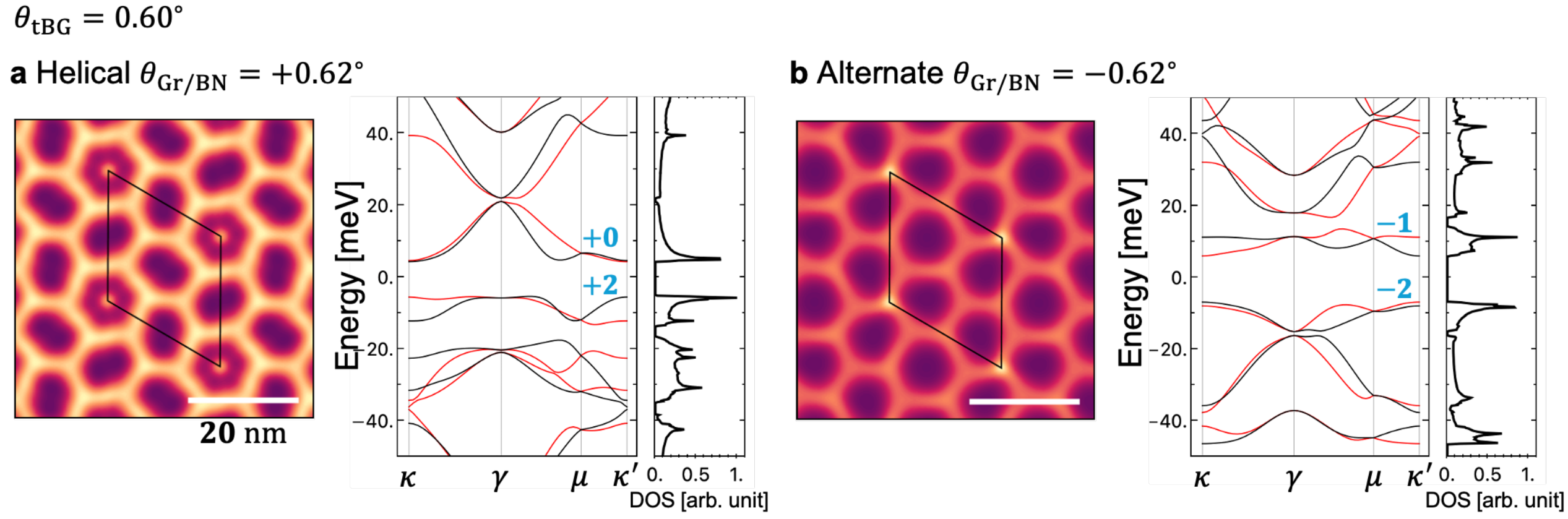

**Figure S7. Optimized lattice structures and electronic band structures for the $2:1$ configuration. a**, Helical twist configuration with $(\theta_{\mathrm{tBG}}, \theta_{\mathrm{Gr/BN}}) = (0.60°, 0.62°)$, and **b**, alternate twist configuration with $(\theta_{\mathrm{tBG}}, \theta_{\mathrm{Gr/BN}}) = (0.60°, -0.62°)$. The black rhombus denotes the unit cell in the lattice structure, and the scale bar corresponds to 20 nm. In the electronic band structures, blue numbers denote the band Chern numbers for the $K$-valley, while black and red lines represent the bands of the $K$ and $K'$- valleys, respectively. The charge-neutral point is taken as the zero energy.

## VIII. Quasi-1D electronic state in the strained configuration

Here, we discuss the electronic properties of the strained commensurate configurations [see Fig. 3e,g,h in the main text] and show that a quasi-one-dimensional band structure emerges. Figure S8 summarizes the optimized lattice structures and corresponding electronic structures for the strained configurations. The top panels show the lattice structures obtained from experiment (left) and theory (right). The middle panels present contour plots of the band structures in momentum space, where the left and right columns correspond to the first valence and first conduction bands measured from charge neutrality, respectively. The bottom panel shows the corresponding density of states (DOS).

As shown in Figure S8b and e, both the helical and alternating configurations exhibit quasi-one-dimensional band dispersion. This behavior originates from anisotropic interlayer coupling between $Gr_1$ and $Gr_2$ induced by strain.

Such quasi-one-dimensional electronic structures have also been reported in strained graphene/h-BN[20] and $WSe_2/MoSe_2$[21] systems. In general, strain in real materials is spatially non-uniform; however, in tBG/h-BN, AA-stacked regions in graphene/h-BN are naturally selected by the moiré pattern, leading to the formation of periodically arranged strained commensurate domains. Therefore, tBG/h-BN provides a robust platform for realizing quasi-one-dimensional electronic states and exploring associated quantum phenomena.

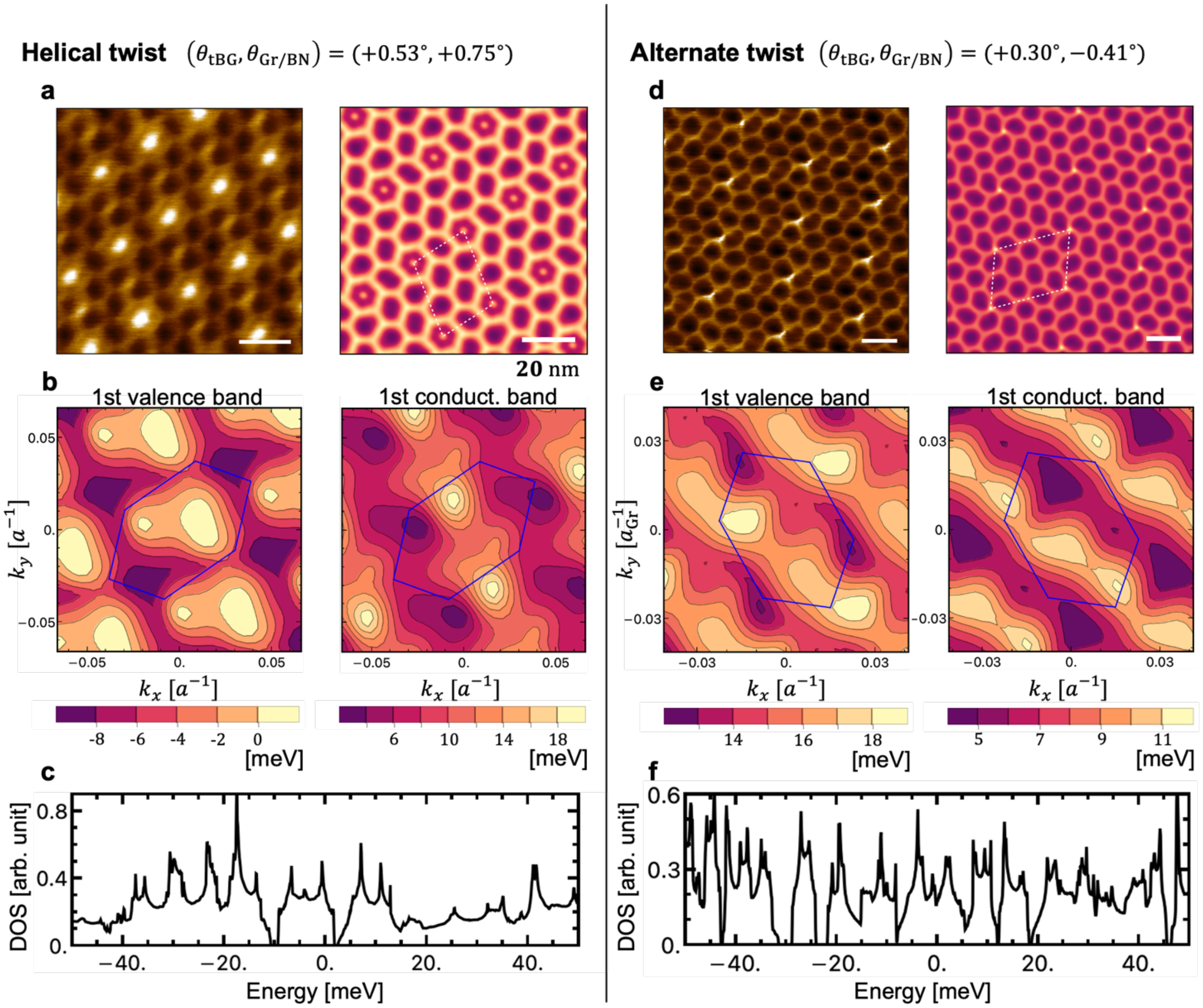

**Figure S8. Quasi-one-dimensional electronic properties in strained commensurate configurations. a**, Lattice structure of the helical twist with $(\theta_{\rm tBG}, \theta_{\rm Gr/BN}) = (0.53°, 0.75°)$ [corresponding to Fig. 3e in the main text], shown as (left) experimentally measured by C-AFM and (right) obtained from continuum-model simulations. Scale bar: 20 nm. **b**, Momentum-space contour plots of the first valence (left) and first conduction (right) bands referenced to the charge-neutral point, showing the $K$-valley band structure. The $K'$ valley is the time-reversal partner of the $K$ valley. The blue distorted hexagon indicates the moiré Brillouin zone. **c**, Density of states as a function of energy. **d–f**, Similar plots for the alternate twist configuration with $(\theta_{\rm tBG}, \theta_{\rm Gr/BN}) = (0.30°, -0.41°)$ [corresponding to Fig. 3h in the main text].

## IX. Experimental conditions

**Table S3. Experimental conditions.**

| | Sample name | Scan area/rate | Contact force | Bias voltage | Top h-BN |
|---|---|---|---|---|---|
| Figures 1d, 3c | Sample 1 | 100 nm/4 Hz | 10 nN | 0.2 V | w/o |
| Figure 1g, 3f | Sample 2 | 150 nm/5 Hz | 5 nN | 0.2 V | w/o |
| Figure 3a | Sample 3 | 200 nm/4 Hz | 10 nN | 0.2 V | w/ |
| Figure 3b, S6c | Sample 4 | 200 nm/4 Hz | 10 nN | 0.2 V | w/ |
| Figure 3d | Sample 1 | 100 nm/4 Hz | 10 nN | 0.25 V | w/o |
| Figure 3e, S4a, S8a | Sample 1 | 100 nm/4 Hz | 10 nN | 0.2 V | w/o |
| Figures 3g, S4e | Sample 2 | 100 nm/6 Hz | 5 nN | 0.2 V | w/o |
| Figure 3h, S4i, S8d | Sample 2 | 150 nm/6 Hz | 10 nN | 0.25 V | w/o |
| Figure 3i | Sample 5 | 100 nm/4 Hz | 10 nN | 0.2 V | w/ |
| Figure 3j, S4m | Sample 5 | 100 nm/4 Hz | 10 nN | 0.2 V | w/ |
| Figure 4a, S6a | Sample 1 | 1000 nm/2 Hz | 10 nN | 0.2 V | w/o |
| Figure 4b, S6b | Sample 2 | 500 nm/1 Hz | 10 nN | 0.2 V | w/o |
| Figure S6d | Sample 3 | 400 nm/2 Hz | 10 nN | 0.2 V | w/ |
| Figure 4c | Sample 1 | 400 nm/4 Hz | 10 nN | 0.1 V | w/o |
| Figure 4d | Sample 1 | 400 nm/4 Hz | 10 nN | 0.1 V | w/o |
| Figure S5 | Sample 1 | 120 nm/4 Hz | 20 nN | 0.3 V | w/o |